\documentclass[11pt]{article}

\usepackage[parsep]{collref}
\usepackage{color}
\pdfoutput=1
\usepackage{amsthm, amssymb,latexsym,amsfonts}
\usepackage{comment}
\usepackage{graphicx}
\usepackage{caption}
\usepackage{amsmath}
\usepackage[usenames,dvipsnames,table]{xcolor}
\usepackage{array}
\usepackage{subcaption}
\usepackage{epstopdf}
\usepackage{enumerate}
\usepackage{tensor}
\usepackage{slashed}
\usepackage[export]{adjustbox}
\usepackage[aligntableaux=center]{ytableau}
\usepackage{dsfont}
\usepackage[utf8]{inputenc}

\newcommand {\be} {\begin {equation}}
\newcommand {\ee} {\end {equation}}
\newcommand {\bes} {\begin {equation*}}
\newcommand {\ees} {\end {equation*}}

\newcommand{\CP}{\mathbb{CP}}
\newcommand{\Z}{\mathbb{Z}}
\newcommand{\N}{\mathbb{N}}

\newcommand{\beq}{\begin{equation}}
\newcommand{\eeq}{\end{equation}}
\newcommand{\ov}{\over}

\def\ie{\begin{equation}\begin{aligned}}
\def\fe{\end{aligned}\end{equation}}
\newcommand{\la}{\langle}

\newcommand{\m}{\mu}

\newcommand{\D}{{\delta}}

\numberwithin{equation}{section}

\def\<{\langle}
\def\>{\rangle}
\def \eps {\epsilon}

\newcommand{\foot}{\footnote}
\newcommand{\ci}{\cite}
\def\ov{\over}
\newcommand{\rf}[1]{(\ref{#1})}
\def\no{\nonumber}

\def\OO{\mc O}
\def \adss {$\text{AdS}_{5}\times S^{5}$\ }

\def\a{{\alpha'}}
\def \ha {{1 \ov 2}}
\def \te {\textstyle}
\def \l {\lambda} 
\def \no {\notag}
\def \iffa {\iffalse} 
\def \ed {\small
\bibliographystyle{ssg}
\bibliography{M2-ext-bib}\end{document}}
   
\def \ZZ  {{\mathbb Z}}
\def \la {\label}
 
\def \l {\lambda} 
\def \gs  {g_{\rm s}}  \def \CP  {{\rm CP}}
\def \str {{\rm s}}
\def \cc {{\rm c}} 
\def \ha {{1\ov 2}}
\def \AA {{\rm A}}
\def \G  {\Gamma} 
\def \te {\textstyle}

\def \r {\rho}  \def \ka {\kappa} \def \s {\sigma} \def \vp {\varphi}
\def \del {\partial} \def \td {\tilde}

\def \V  {{\rm V}}
\def \ha {{1 \ov 2}}
\def \b {\beta}
\def \sql {{\sqrt{\l}}\ }
\def \del{\partial}
\def \a {\alpha}

\def\ov{\over}
\def \ci {\cite}

\def \foot {\footnote}

\def\la{\label} 

\def \cD {{\cal D}} 
\def \no {\nonumber}
\def \adss {AdS$_5 \times S^5\ $}
\def \a {\alpha } 
\def \eps {\epsilon}

\def \V  {{\rm V}}

\def \bea {\begin{align}}
\def \eea {\end{align}}
\def \T {{\rm T}}
\def \ha {\tfrac{1}{2}}
 \def \rr  {{\rm p}}
 \def \k {{\rm p}}

\def \ed {\small  
\bibliographystyle{JHEP-v2.9}
\bibliography{M2-ext-bib2}
 \end{document} }
 
\def \we {\wedge} \def \OO {{\cal O}}
\def \a  {\alpha}
 
\def \td {\tilde}
 
 \def \abjm {_{\rm ABJM}}
\def \bl {{\bar \l}}

\def \adsc {AdS$_4\times \CP^3$}  
\def \adsz  {AdS$_{4}\times S^{7}/\mathbb{Z}_{k}$ }
\def \ba {\begin{align}} 
\def \ads {{\rm AdS_4}}
\def \LL {{\rm L}}
\def \D  {\Delta}  \def \N  {{\cal N}}
\def \adsm {$\text{AdS}_{4}\times S^{7}/\ZZ_k$\ }
\def \ff {{\rm f}}   \def \cor {  } 
  \def \np {\newpage}
  \def \brf {Brehmstrahlung function\ }
  \def \lp {\ell_P}
\def \ba {\begin{align}}  \def \rZ  {{\rm Z}} \def \Ze {{\cal Z}}
   
\def \tr {{\rm tr}}
\def \T  {{\rm T}} \def \four {\tfrac{1}{4}} 
   
\def \rT {\T}
\def\lon {``long"\ }  \def \sho {``short"\ }
\def \EE {{\cal Q}}
\def \s {\xi}
\def \bz {\bar z}
 \def \E {{\cal E}}
\def \ff {{\rm f}}\def \sql {\sqrt{\l}}\def \rT  {{\rm T}}
 \def \g {\gamma} 
 \def \J {{\cal J}} \def \sqbl {\sqrt{\bl}}  \def \E {{\cal E}} 
 \def \hE {\hat \E}  \def \cG {{\cal G}}

\def \k {\ka} \def \Z  {{\cal Z}}
\def \nn {l}

\def\adssZ{$\text{AdS}_{4}\times S^7/\mathbb{Z}_k \ $}
\def\ads{$\text{AdS}_{4}$}

\def \m2{$\text{M2}$}
\def \adsc {AdS$_4\times \CP^3$} 

\def \y {\vp} \def \g {\gamma}
\def \bc {{\rm c}}
\def \OD {{\rm D}}
\def \om {\omega} 
\def \rod {R_{11}}
\def \rw {{\rm w}}
\def \cD {{\cal D}}
\def \rn {{\rm n}}  \def \NN {{\rm X}}  \def \rK {{\rm K}} \def \rA {{\rm A}}
\def \rF {{\rm F}}
\def \qq  {{\rm q}}
\def \rr {{\rm w}} 
\def \sym {{\rm SYM}} \def \abjm {_{\rm ABJM}}

\begin{document}
\vspace{ -3cm} \thispagestyle{empty} \vspace{-1cm}
\begin{flushright} 
\end{flushright}
 \vspace{-1cm}
\begin{flushright}  Imperial-TP-AT-2024-03\\ 
  \date {  } \today 
\end{flushright}
\begin{center}
 \vspace{1.2cm}
{\Large\bf
Non-planar corrections in ABJM theory from quantum M2 branes}
 
 \vspace{0.8cm} {
 Simone Giombi$^a$, Stefan A. Kurlyand$^{b}$  and Arkady A. Tseytlin$^{b,}$\footnote{ Also  at 
the  Institute  for Theoretical and Mathematical Physics (ITMP)  and Lebedev Institute.}
    }\\
 \vskip  0.5cm

{\em
$^{a}$Department of Physics, Princeton University, Princeton, NJ 08544, USA \\
\vskip  0.1cm
  $^{b}$Theoretical Physics Group, Blackett Laboratory,  Imperial College London,
 SW7 2AZ, U.K.
  }

\normalsize
\end{center}

 \vskip 1.2cm

 \begin{abstract}
 The quantization of semiclassical  strings  in AdS spacetimes 
yields predictions  for  the strong-coupling behaviour of the scaling dimensions of  the corresponding operators in the planar limit of the dual  
gauge  theory.
 Finding  non-planar corrections  requires  computing  string loops (corresponding to torus and higher genus surfaces), which is a challenging
 task.  It turns out  that 
  in the case of the $U_k(N) \times U_{-k}(N)$ 
  ABJM theory there is an alternative approach: one may semiclassically quantize M2 branes 
  in AdS$_{4}\times S^7/\mathbb{Z}_{k}$
  which are wrapped around the 11d circle of radius $1/k= \l/N$. Such M2 branes are the M-theory generalization of the strings in  AdS$_4\times \CP^3$. 
In this work, we show that by expanding in large M2 brane tension \( T_2 \sim \sqrt{kN} \) for fixed \(k\), followed by an expansion in large \(k\), we can predict the large \(\lambda\) asymptotics of the 
non-planar 
corrections to the dimensions of the dual ABJM operators.
As a specific example, we consider the   M2 brane configuration that generalizes the long folded string with large spin in AdS\(_4\), and
compute the 1-loop correction to its energy. This calculation allows us to determine 
non-planar corrections  to the universal scaling function or cusp anomalous dimension.
We extend our analysis to the semiclassical M2 branes that generalize the ``short" and ``long" circular strings with two equal angular momenta in \(\rm{CP}^3\). The ``short" M2 brane  corresponds to a dual operator whose dimension at strong coupling scales as \(\Delta \sim \lambda^{1/4} + \dots\), and we derive the leading non-planar correction to it.


\end{abstract}


\newpage
\tableofcontents

\renewcommand{\theequation}{1.\arabic{equation}}
 \setcounter{equation}{0}
\setcounter{footnote}{0}

\normalsize
\section{Introduction and summary}

One of the   challenging problems in
 superconformal quantum  field  theories  like $\N=4$ SYM and ABJM  \ci{Aharony:2008ug} 
 ones, which admit a
  large $N$  expansion, is to compute the conformal 
 dimensions $\Delta$ of primary operators as functions of the 't Hooft coupling $\l$  and $N$.
In general, 
\be  \Delta (\l,N) = \D_0(\l) + {1\ov N^2}\D_1(\l) + {1\ov N^4}\D_2(\l) + ...  \ . \la{01} \ee
Here the planar  part  $\D_0(\l) $ is controlled by integrability, 
 and, expanded at  large $\l$,  it can be  matched to  the large tension expansion of string energies in the
  dual string theory (see, e.g.,  \ci{Beisert:2006ez,Basso:2007wd,Beisert:2010jr}).
  Little, however,  is known   about  the  explicit form of  the  non-planar   correction
  $\D_1$, $\D_2$, \ldots . 
  In the $\N=4$ SYM theory 
   the  first non-planar correction to the cusp  anomalous dimension $f(\l,N)$ 
  appearing  in  the large spin  expansion of the   dimension of   an  operator like $\OO= \tr (  \Phi  D_+^S \Phi ) $
  \be \Delta\big|_{S\gg 1}  = S + f(\l,N) \log S + ... \la{1200}
  \ee  
    starts  at 4-loop order in  the weak coupling  expansion  \ci{Henn:2019swt} (see also \ci{Boels:2017skl,Henn:2019rmi})
  \begin{align} \la{508}
&  f(\l,N) =\tfrac{1}{ (2\pi)^2} \Big[\l -  \tfrac{1}{48} \l^2 +  \tfrac{11}{11520} \l^3 
-   \big( c_{4} + {d_4\ov N^2}     \big)\l^4  + \OO(\l^5)\Big],\ \ 
  \\
  & 
  c_{4}= \tfrac{73}{20160\times 64} \ + \tfrac{1}{8(2\pi)^6} \zeta^2(3) \ , \qquad 
  \qquad  d_{4}= \tfrac{31}{5040\times 64}  + \tfrac{9}{4(2\pi)^6 } \zeta^2(3) \ .  \la{509}
  \end{align}
 The   $\l^4\ov N^2$  term   appears to  be   universal --  
  it  is the same for  any matter  content   \ci{Korchemsky:2017ttd}  as it originates 
    from the quartic Casimir of $SU(N)$. This   suggests
    that  in  all anomalous dimensions  computed at weak coupling the $1/N^2$ correction should first appear  at  4 loops, i.e.
    $\Delta_1$ in \rf{01}   
    should be    given by  
    \be \la{02}  \  \D_1\big|_{\l \ll 1}  = d_4 \l^4 + d_6 \l^6 +  ... \ . \ee
Indeed,    
   similar non-planar behaviour is found  for  the anomalous  dimensions of twist-2 operators  with general Lorentz spin \ci{Kniehl:2020rip,Kniehl:2021ysp}\foot{According to \ci{Kniehl:2021ysp}  at large spin at weak   coupling we should expect 
   the anomalous dimension
    depending on   spin  as  
  $\D_1(S)=  {\l^4} \big(d_4  \log S + e_1 + { 1\ov S}  e_2 + ...  \big)$, 
    where $e_1,e_2 $, like $d_4$,  are  given by combinations of $\zeta$-values. 
 We thank V. Velizhanin  for a  comment on this expansion
   and informing  us that the coefficient of the $ { 1\ov S} \log S $  term 
   happens  to be  zero.
   }  
   and also for the 
    Konishi   operator  \ci{Velizhanin:2009gv,Velizhanin:2010ey}
   where $d_4 \sim \zeta(5) $   (see also    \ci{Fleury:2019ydf}).
 
 Less is known  about non-planar corrections  in the case of the ABJM theory.\foot{Study of  non-planar corrections at  leading order at weak coupling in the ABJM theory was initiated in 
  \ci{Kristjansen:2008ib}. 
  In \ci{Griguolo:2012iq} the      2-loop correction to the  cusp anomaly was found not to contain a non-planar part,
  but  in   sect 4.1 \ci{Bianchi:2013iha} (cf.  also \ci{Henn:2023pkc}) the opposite was claimed.
  We thank M. Lagares  for pointing this out.}
Given a close  analogy   with the $\N=4$ SYM theory,  it is   natural to expect that  here the first non-planar 
correction  should   also   appear at 4-loop order  as in  \rf{02}. 
 
 One may conjecture that it may be possible to compute 
 $\D_1(\l) $ in  \rf{02}  to all orders  utilizing somehow the 
 integrability of the planar theory (cf., e.g.,  \ci{Kristjansen:2010kg,Bargheer:2017nne,McLoughlin:2020siu}). 
 If one could do this,  then expanding the exact  expression for $\D_1(\l) $ at strong coupling,  
 $\D_1\big|_{\l \gg 1} \sim  \l^p + ...  $, 
 one would then   determine   the power $p$  of the leading  term.
 It    could then   be  compared to the  dual   string theory side  where 
     finding  the leading non-planar correction  requires  computing string 1-loop (torus)  correction      to  string energies,   
     a complicated  open problem. 

Remarkably,   as we shall demonstrate  below, there is a way to  find  
the   strong-coupling asymptotics  of the non-planar  corrections 
$\D_1(\l), \D_2(\l),\ldots$  in  the case of the 
ABJM 
model  using its duality  to M-theory or theory of quantum  M2 branes.
It turns out that  a
  semiclassical M2  brane quantization in \adsm    captures the 
      leading order $\a'\sim {1\ov \sqrt \l} $  terms at each order in  the string coupling $\gs^2\sim {1\ov N^2}$  expansion.

In particular, we will show  that for the ABJM cusp anomalous dimension   the  strong-coupling  scaling 
of the leading non-planar correction is ${\l^2\ov N^2}$.
In general, the prediction for the   structure of the  large  $\l$   expansion  of   the 
$1/N^{2s}$ coefficients   in \rf{1200} is  
 \begin{align}  \la{03}
&f(\l, N) = f_0(\l) + {1\ov N^2} \ff_1(\l) +   {1\ov N^4} \ff_2(\l)+  ...\ , \ \ \qquad   f_0\big|_{\l \gg1} = \sqrt{2\l} + \ff_0(\l) \ , 
\\
& \qquad \ff_s (\l)\big|_{\l \gg1} =  
\te \l^{2s} \big(a_{1s}+  {1\ov \sql} a_{2s} + ... \big)\ , \ \ \ \ \ \  \ \ \  \ \ \  s=0, 1,2, ...
\ . \la{04}
 \end{align}
 The few   leading  coefficients in the strong-coupling expansion of the planar part $f_0(\l)$ 
 can be  found  (as in the \adss case 
 \ci{Gubser:2002tv, Frolov:2002av,Roiban:2007jf,Roiban:2007dq,Giombi:2009gd})
  by quantizing 
  the  long folded spinning string  in \adsc \ci{McLoughlin:2008ms,Alday:2008ut,
 Krishnan:2008zs,Bianchi:2014ada}.
 The coefficients $a_{1s}$ of the leading non-planar   contributions 
  will be computed below  from  the 1-loop  3d world-volume 
   correction to the energy of a  semiclassical   M2 brane   spinning in AdS$_4$  and 
    wrapped on the 11d circle in $S^7/\ZZ_k$  of radius ${1\ov k} = {\l\ov N}$.
  The  subleading  $a_{2s} $  coefficients  may be 
     determined from   the 2-loop M2  brane correction, etc. 
      
   \subsection{Semiclassical expansion for M2 brane  in \adsz} 
   
While the M2 brane   action   \ci{Bergshoeff:1987cm,Bergshoeff:1987qx}  
 is formally non-renormalizable, 
the  semiclassical expansion of the   corresponding 
  path integral near a  ``minimal-volume"  solution  with a non-degenerate induced 
  3d metric is well defined (at least at the 1-loop order 
   \cite{Kikkawa:1986dm,Duff:1987cs,Bergshoeff:1987qx,Fujikawa:1987av,Mezincescu:1987kj,Forste:1999yj}
where   there is  no  logarithmic UV   divergences  in a  3d theory). 
  Recent work 
\cite{Drukker:2020swu,Giombi:2023vzu,Beccaria:2023ujc,Beccaria:2023sph,Drukker:2023jxp,Drukker:2023bip}
provided  a  convincing evidence  that the semiclassical  quantization of  the 
M2  brane is indeed consistent. It was 
demonstrated that   1-loop    M2  brane  corrections in \adsz  
and     AdS$_7 \times S^4$   match  the dual  gauge theory (localization) 
predictions for  several  ``supersymmetric"  observables 
 --  
defect anomaly,   $\ha$ BPS  Wilson loop  and 
 instanton contributions to the supersymmetric partition function  (superconformal  index)  
in the 3d ABJM and  also   6d (2,0) theory. 

This  provides a motivation   to apply  similar semiclassical M2  brane  quantization approach also 
to  ``non-supersymmetric"  observables   like  non-planar corrections to ABJM  anomalous dimensions 
that  are not controlled  by integrability or localization.


Let us   briefly review  some  basic relations and notation that we will use below. 
The $U_k(N) \times U_{-k}(N)$   ABJM theory  expanded  at large $N$  for  fixed $k$  is dual to M-theory 
on 
AdS$_4 \times S^7/\ZZ_k$  background 
with the metric and 3-form given by 
\begin{align} \la{05}
&\qquad \qquad ds_{11}^{2} = \LL^{2} \Big(\tfrac{1}{4}ds^{2}_{\rm AdS_{4}}+ds^{2}_{S^7/\ZZ_k}\Big),\qquad \ \qquad 
  {\LL} = (2^{5}\pi^{2} N k)^{1/6} {\ell_{P}} \ ,  \\
  & \qquad \qquad ds^2_{{\rm AdS}_4}  = - \cosh^2 \r\, dt^2 + d \r^2 + \sinh^2 \r\, ( d\a^2  + \cos^2 \a \, d \b^2) \la{100s},  \\
  &\qquad \qquad  ds^{2}_{S^7/\ZZ_k} =ds^{2}_{\rm CP^{3}} +   {1\ov k^2} (d\vp +   k {\rm A})^{2}\, , \qquad  
  \ \ \ \  \vp\equiv  \vp + {2 \pi } \ , \la{06}\\
  &\qquad \qquad  
  ds^2_{{\rm CP}^3}=d z^a d \bz_a -  \bz_a  z^b d z^a   d \bz_b \ , \qquad 
  {\rm A}=\te {1\ov 2 i}( \bz_a d z^a - z^a d \bz_a) \ , \ \ \ \   \bz_a z^a=1, \ \ \   a=1,...,4
    \ , \la{07}\\
  & \qquad \qquad C_3 = - \te {3\ov 8}  \LL^3 \cosh \r\,   \sinh^2 \r\,   \sin  \a\, dt \we  d\r  \we  d \b \   \la{7} \ . 
  \end{align} 
  Taking also $k$ large   with $\l\equiv { N\ov k}$  fixed  corresponds to the  't Hooft expansion of the 3d gauge theory in which it is dual to the perturbative type IIA  string theory in AdS$_4 \times \CP^3$ with the   coupling $\gs$ and the 
  effective   dimensionless  string  tension $\T$  defined with respect to the  radius of the AdS$_4$ part
    given by 
     (we set $\ell_P=\sqrt{\a'}$ as in Appendix A in \ci{Beccaria:2023hhi})
     \begin{align}
&ds_{10}^{2} = L^{2} \Big(\tfrac{1}{4}ds^{2}_{\rm AdS_{4}}+ds^{2}_{\rm CP^{3}}\Big),\qquad\qquad  \ \ L = \gs^{1/3}\,\LL \ , \la{08}\\
& \gs =  \big(\frac{\LL}{k\, \ell_{P}}\big)^{3/2} = \frac{\sqrt\pi\, (2\l)^{5/4}}{N} \ , \ \ \ \ \ \ \  \l= { N \ov k } \ , 
\qquad \T = \frac{\frac{1}{4} L^{2}}{2\pi \alpha'} 
= \sqrt{\l\ov 2}= {\sqrt{\bl} \ov 2 \pi} \  , \ \ \ \  \ \  \bl \equiv 2\pi^2 \l \ , \la{09} \\
&\qquad\qquad    {{1\ov k^{2}} ={\l^2 \ov N^2}=   \cor {{ \frac{\gs^2 }{8\pi\, \T } } }}  \la{010}
\ . 
\end{align}
The M-theory expansion  corresponds to ${\LL \ov \lp}\gg 1 $   or     large $N$  for  fixed $k$, i.e. 
the expansion in the large  effective dimensionless M2  brane tension
\be \la{011} 
\T_2\equiv \LL^3  T_2= {1\ov \pi} \sqrt{ 2 N k } \ , \ \ \  \qquad \qquad  T_{2} = \frac{1}{(2\pi)^{2}\ell_{P}^{3}} \ . 
\ee 
Here  $\T_2$ is defined with respect to the radius of $S^7/\ZZ_k$ part so that  it is related to the string tension  in \rf{09} 
as  (note also that  in general ${1\ov k} {\LL^3 \ov \ell_{P}^{3}} = {L^2 \ov \a'}$)
\be  \T = {1\ov 4}  {2\pi\ov k} \T_2 \ .\la{0115}
\ee
The observables   that can be computed in the semiclassical M2  brane expansion  can be written as 
\be 
F = \T_2 F_0(k)+ F_1 (k)  +  ( \T_2)^{-1}  F_2(k) + ... \ , \qquad \qquad \T_2 \gg 1 \ .  \la{012} 
\ee
This   corresponds to the large $N$  expansion for fixed $k$. 
Expanding \rf{012}  further at large $k$,   it may be rewritten as a  large $N$ expansion for fixed $\l= {N\ov k}$ 
or string-theory expansion in  $\gs$ for fixed  $\T= \sqrt{\l\ov 2}$.

Below we will assume that dimensions  of ABJM operators  with large quantum numbers
that, in the planar expansion,  are dual to semiclassical strings in AdS$_4 \times \CP^3$, 
 may be computed  
as AdS$_4$ energies of semiclassical M2  branes in 
 \adsm   that are wrapped on the  11d circle $\vp$ in \rf{06}.
 They  will thus  have topology $\Sigma^2 \times S^1$, i.e.   will    generalize 
 the corresponding string solutions reducing to them 
upon  the ``double dimensional reduction" \ci{Duff:1987bx}.

Given a M2  brane solution with a  non-degenerate induced 
3d metric it is 
straightforward to   expand  the corresponding path integral  at large $\T_2$  (using, e.g., a   static gauge as in 
 \cite{Drukker:2020swu,Giombi:2023vzu,Beccaria:2023ujc}).
The M2  brane action  is 
  \begin{align} \la{013}  &S= S_B + S_F\ , \qquad \ \ S_B = S_\V+S_{\rm WZ},  \qquad \ \ 
S_\V =- T_{2}\, \int d^{3}\s\, \sqrt{ -g}\ ,   
\\ &
S_{\rm WZ} = -\, T_{2}\, \int d^{3}\s\, \tfrac{1}{3!}\eps^{ijk}C_{MNK}(X)\, \partial_{i}X^{M}\partial_{j}X^{N}
\partial_{k}X^{K},  \la{032} \\
&  S_F = 
T_2  \int d^3\s\Big[ \sqrt{- g} \,   g^{ij}   \del_i X^M  \,   \bar \theta \,  \Gamma_M \hat D_j \theta 
 - \tfrac{1}{2}\eps^{ijk}  \del_i X^M  \del_j X^N  \,  \bar \theta \,  \Gamma_{MN}  \hat D_k \theta 
  + ...\Big]\ , 
  \la{33} \\ 
 &  g_{ij} = \del_i X^M  \del_j X^N   G_{MN} (X), \qquad \ \  \
   G_{MN} = E^A_M E^A_N, \qquad  \ \ \
  \Gamma_M  =   E_M^A(X) \Gamma_A\  \la{014}\ ,    \\  
  & \hat D_i = \del_i X^M   \hat D_M , \ \qquad 
  \hat D_M = \del_M + \four \G_{AB} \Omega^{AB}_M    - \tfrac{1}{288} (\G^{PNKL}_{\ \ \ \ \ \   \ M} - 8  \G^{PNK}\delta^L_M ) F_{PNKL} \  . \la{034}
\end{align}
The leading  classical and 1-loop  contributions to  the (euclidean)  M2 brane  partition function may be written as 
\begin{align} \la{036} &
\rZ_{\rm M2} =\int [dX\, d\theta]\, e^{- S[X,\theta]} = \  \Ze_1\,  e^{-\T_2 \bar S_{\rm cl}}\big[1 + \OO (\T_2^{-1} )\big]   \ , \ \ \ \\
&\Ze_1 = e^{-\G_1} \ , \ \ \ \qquad \   \G_1 =\tfrac{1}{2} \sum_i  \nu_i  \log \det \OO_i \ , 
\ \la{037} \end{align}
where $\bar S_{\rm cl}$  and fluctuation operators $\OO_i $ may  depend   on the parameter  $k$  or the inverse 
radius of the 11d circle   in  the 11d metric 
\rf{05},\rf{06}  and other  parameters  of a given classical solution (like rotation frequencies, winding numbers, etc.). 
Then  $F=-\log \rZ_{\rm M2}$ will have the   form   given   in \rf{012}. 

 It is important to note that  
   we are expanding  near  just one M2 brane saddle  (i.e. we are not summing over 3d topologies).
 Interpreted  from the string theory point of view, this world-volume loop  expansion already captures  contributions of  all higher string loops (as well as  the dependence on the string tension).
  Indeed, the  classical M2 brane   action   encodes the dependence on the  string coupling $\gs$  (cf. \rf{010}) 
  via  its   dependence on  the parameter 
  $k$ of the 11d background metric \rf{05},\rf{06}  in which the M2 brane  is embedded.

  Since   the membrane is assumed to  have   $\Sigma^2 \times S^1$   topology,   
  in the static gauge the $S^1$ direction  may be  identified with 11d direction $\vp$ in \rf{06}.
  Then expanding all  3d world volume fields  in Fourier modes in the  $S^1$  coordinate   the M2  brane action 
  may be written as a  2d action for   the ``massless" 
  2d fields, representing the  corresponding type IIA string action in AdS$_4 \times \CP^3$,   
  interacting with an infinite set    (``KK tower")    of   2d fields 
  with  masses 
$m^2_l = {l^2\ov R^2_{11}} = l^2  k^2 = l^2 {8\pi \T\ov \gs^2} $  $(l=1,2,...)$
depending on  the string coupling. 

These  massive 2d fields decouple only in the strict $\gs\to 0$ limit, 
while  in general their   contributions   will  encode the   string loop corrections to \rf{037} or \rf{012}. 
Integrating them    out  one would get an effective non-local  action for the ``massless"  (string or $l=0$ level) modes.  
That would realise the idea of having an effective  string action on 2-sphere supplemented 
by  ``handle operator"  contributions  that account  for the usual  sum  over the  2d topologies 
(cf.  \ci{Tseytlin:1990vf, Skliros:2019bqr,Seibold:2024oyr}).

\subsection{Example:   $1/N$ expansion of $\ha$ BPS Wilson loop}

To illustrate the above discussion let us review  the  case   of the $1/N$    expansion   
of the $\ha$ BPS Wilson loop expectation value in the ABJM theory  that can be  reproduced 
 by the   semiclassical M2  brane computation,  
as was   demonstrated in  \ci{Giombi:2023vzu}.
 The exact (at large $N$  and for $k>2$)  result  found  from localization matrix model on the gauge  theory side
 is \ci{Klemm:2012ii} 
 \begin{equation}
 \langle W \rangle= \frac{1}{2\, \sin \frac{2\pi}{k} }\ \frac{{\rm Ai}\left[( {\pi^2\ov 2} k)^{{1}/{3}}\left(N-\frac{k}{24}-\frac{7}{3k}\right)\right]}{{\rm Ai}\left[( {\pi^2\ov 2} k)^{{1}/{3}}\left(N-\frac{k}{24}-\frac{1}{3k}\right)\right]}\ . 
\label{2190}
\end{equation}
Expanded in  large $N$  at fixed $k$   and then further   in large $k= {N\ov \l}$  this  may be written as 
 \begin{equation}
\langle W\rangle = \frac{1}{2\sin\frac{2\pi}{k}}\ e^{\pi  \sqrt{\frac{2N}{k}}}\Big[
1-\frac{\pi  \left(k^2+32 \right)}{24 \sqrt{2} \, k^{3/2}}\frac{1}{\sqrt{N}}+\OO(\frac{1}{N})
\Big] = \frac{1}{2\sin \frac{2\pi\l}{N}}\  e^{\pi  \sqrt{{2\l}}}\Big[
1-\frac{\pi }{24 \sqrt{2}}\frac{1}{\sqrt{\l }}+\OO(\frac{1}{N})
\Big] .
\label{350}
\end{equation}
The first expansion  here   may be  written  as a semiclassical expansion  for large effective M2 brane   tension  $\T_2$ in \rf{011} 
\begin{equation}
\langle W\rangle  =  \frac{1}{2\sin \frac{2\pi}{k} }\, e^{ {\pi^2 \ov k} \T_2}
\Big[
1-\frac{k^2+32 }{24k }({\T_2  })^{-1}+\OO\big(( \T_2)^{-2}\big)
\Big] \ ,  \la{017}
\end{equation}
so that it takes the form of \rf{036}   or, equivalently,    $\log \langle W\rangle $ takes the form of \rf{012}
\be \la{0177}
\log \langle W\rangle = {\pi^2 \ov k} \T_2  - \log\big( 2\sin \frac{2\pi}{k}  \big)
-\frac{k^2+32 }{24k }({\T_2  })^{-1}+\OO\big(( \T_2)^{-2}\big)\ . 
\ee
The exponential  factor in  \rf{017} comes from the  value of the  action of the  M2  brane 
 wrapped  on AdS$_2 \times S^1$ 
(ending on a circle at the boundary of AdS$_4$).  
 The prefactor $\frac{1}{2\sin \frac{2\pi}{k} }$  was  reproduced in \ci{Giombi:2023vzu}   as 
 the   1-loop contribution \rf{037} of the corresponding 3d  fluctuations.  
 The subleading $(\T_2)^{-1}$   term in \rf{017}  should originate from the  2-loop  M2  brane  contribution, etc. 
 
Expressed in terms of  the string tension and the string coupling in \rf{09}  the   prefactor in \rf{017} may be written as \ci{Beccaria:2021ksw} (cf. \rf{010})
\be 
  \langle W\rangle
 = {1  \ov 2  \sin \big( \sqrt\frac{\pi}{2}\,\frac{\gs}{\sqrt \T}\big)   } \, e^{2\pi\,\T}  \Big[1+ O(\T^{-1})\Big] 
=   {\sqrt \T \ov\sqrt{2 \pi} \gs }\, e^{2\pi\,T} \Big[1 
  +{\pi \ov 12}   { g^2_\str \ov \T } +{7  \pi^2\ov 1440}   \big({ g^2_\str \ov \T }\big)^2 +...\Big]
 \Big[1+ O(\T^{-1})\Big] 
 \ . \la{0127} \ee
Thus the large $k$ expansion of the 1-loop M2 brane factor 
$\frac{1}{2\sin \frac{2\pi}{k} }$   captures the  leading  large string tension (or large $\l$)  corrections  at each order  in $\gs^2$,  while  the 2-loop  and higher M2  brane corrections  determine the subleading in $\T^{-1}\sim {1\ov \sql}$ terms at each order in $\gs^2$. Equivalently, \rf{017} or 
\rf{0127}   implies that 
 \be \la{0128}
\langle  W \rangle
= \   {\sqrt \T \ov \gs } e^{2\pi \T} \Big[  \big(  \cc_{00}
 + {\cc_{10}\ov \T}+ ...\big)
  + { g^2_\str \ov \T } \big(  \cc_{01}  +{\cc_{11}\ov \T}+ ... \big)
+    \big({g^2_\str \ov \T } \big)^2 \big( \cc_{02}    
 +{\cc_{12}\ov \T}+... \big)+ ... \Big]\ ,  \ee
where  $\cc_{0r}$  ($r=0,1,2,...$)  are determined  by the 1-loop M2  brane contribution, 
$\cc_{1r}$ -- by the 2-loop M2  brane contribution, etc. 
From the  perturbative \adsc\  string theory   perspective, the 
$\cc_{0r}$ coefficients represent the leading in  $\T^{-1}\sim \a'$  terms  at each  order in the string loop (genus) expansion, i.e. $\cc_{00}$ is the 1-loop (in string world sheet  sense) 
coefficient on the disk,  $\cc_{01}$  is its counterpart  on the disk with one handle, etc. 

\subsection{Cusp  anomalous dimension from fast-spinning M2 brane in AdS$_4$}

The same pattern of non-planar corrections  should  apply   also to other observables 
like  anomalous dimensions,  and, in particular, to the  cusp anomalous  dimension.
Namely,  the semiclassical quantization of the M2  brane   generalization of the long spinning  folded  string in AdS$_4$  should lead to the following  expansion  for 
$f(k,\T_2)$   in the M2 brane energy or dimension  \rf{1200}  (cf. \rf{012}) 
\begin{align} \la{0129}
f (k, \T_2) =&  {\pi\ov k} \rT_2 
+ q_0(k)  + {q_1(k)( \rT_2)^{-1}} + {q_2(k)(\rT_2)^{-2}}  + ...\ , \\
q_{r}(k) =  &k^{r} \Big( p_{r}^{(0)} +   {p_{r}^{(1)}\ov k^{2}} + {p_{r}^{(2)}\ov k^{4}}  + ...\Big)\ , \qquad \ \ r=0,1,2, ... \ .  \la{0132}
\end{align}
Since $\T_2 \sim k \sql$ (cf. \rf{011})
   the specific  large $k$  asymptotics of $q_{r}(k)$ in \rf{0132} is the one required 
to match the  inverse string tension  ($ {1\ov  \sql}$) expansion in the strict  tree-level (planar)  string theory
limit or  
 to match the  structure of the expected  $1/N^2$ expansion in the 't Hooft limit  on the gauge theory   side
(cf. \rf{03},\rf{04}). 

Thus
the condition of matching the string theory expansion  like \rf{0128} 
 fixes  the structure of the large $k$ terms 
 in  the  coefficient functions in the general expression for the semiclassical expansion
in  \rf{012}. 
This   requirement  assumes 
that the ``double dimensional reduction"  relation between  the M2 theory and string theory observed  at the  classical   action 
level extends also to the quantum level. This is  
 implied  by  the structure of the  M2 action as an effective 2d  action
containing massive KK modes in $S^1$ direction  which should decouple in the $k\to \infty$ limit  (assuming that  the theory turns out  to be 
  well defined in the UV). 
As in the Wilson loop  case reviewed  above,  we   will  explicitly verify this 
at the  1-loop $q_0(k)$  level below. 

In particular, the leading  large $\l$  asymptotics of each term in the  $1/N^{2}$ expansion, 
 i.e. the  coefficients $a_{1s}$ in \rf{04},   should be  the same as 
  the coefficients $p_{0}^{(s)}$ in the large $k$
 expansion of the 1-loop  M2  brane  function $q_0(k)$, i.e. 
 \be \la{045}
 q_{0}(k) =   p_{0}^{(0)} + \bar q_0(k) \ , \qquad \bar q_0(k)=  {p_{0}^{(1)}\ov k^{2}} + {p_{0}^{(2)}\ov k^{4}}  + ...\ , \qquad \ \ \ 
 a_{1 s}=   p_{0}^{(s)}\ , \ \ \ \ \ \   s=1,2,...
 \ . 
 \ee
 As  we  will find  below (cf. \rf{010})
 \be \la{777}
p_{0}^{(0)}= - {5 \log 2\ov 2 \pi} \ , \qquad 
\ \ \   \bar q_0=   \frac{2 \pi }{3 k^2} +\frac{2 \pi ^3}{45 k^4 }   + ...
  =   { \gs^2\ov 12\T} +  { \pi \gs^2\ov  1440  \T^2}  + ...
 =   \frac{2 \pi \l^2  }{3 N^2 } +\frac{2 \pi ^3\l^4 }{45 N^4 }   + ...
 \ . \ee
At the  same time, 
the  string world-sheet loop corrections   that  represent  the 
$1 \ov \sql$  expansion of the planar function  $\ff_0$ in \rf{03},\rf{04} 
 come from the  leading large $k$ term in the  $q_r(k)$ functions (with coefficient $p_{r}^{(0)}$)
   in \rf{0129} or  explicitly (cf. \rf{011}) 
 \be\la{044}
 \ff_0(\l)\big|_{\l \gg1} =  a_{10}+  {1\ov \sql} a_{20} +{1\ov (\sql)^2} a_{30} +  ... \ , \ \ \ \ \ \ \ \ \ \ 
  a_{r+1, 0}=  ( {\pi\ov \sqrt 2})^{r} \,  p_{r}^{(0)}\ , \ \ \ \ \ \   r=0,1,2,...
 \ . 
 \ee
To recall,  by  direct perturbative large  tension expansion for a  long folded  spinning    string in AdS$_4 \times \CP^3$ one finds  that at  the 1-loop   \ci{McLoughlin:2008ms,Alday:2008ut,
 Krishnan:2008zs}  and 2-loop \ci{Bianchi:2014ada} orders 
 \begin{align} 
f_0(\lambda) &= \sqrt{2\lambda }  + \ff_0(\l) =\sqrt{2\lambda } 
  - \frac{5 \log 2}{2 \pi } - \big(\frac{K}{4\pi^2} + 
\frac{1}{24}\big)\frac{1}{\sqrt{2\lambda}} + {\cal O}({1\ov (\sqrt{\lambda})^{2} })  \la{555}\ ,
\end{align}
where  $K$ is the Catalan's constant.
Expressed in terms of the ``renormalized tension''   containing  a $\log 2$   correction \ci{McLoughlin:2008he}  
\be \la{88}
 h(\lambda)\big|_{\l \gg 1} = \sqrt{\frac{\lambda}{2}} - \frac{\log2}{2\pi} - \frac{1}{48\sqrt{2\lambda}} 
+ ... \ ,   
\ee
eq.\rf{555}  takes the same  form as in the \adss (i.e.  $\N=4$ SYM)  case 
\begin{align}
f_0(\lambda) &=  2 h(\l) 
 - \frac{3 \log 2}{2 \pi } - 
\frac{K}{8\pi^2\,}  h^{-1}(\l) + {\cal O}\big(h^{-2}(\l)\big)\,. \la{155}
\end{align}
In general, the ABJM cusp anomalous dimension is  expressed in terms  of the SYM one  as ~\ci{Gromov:2008qe}:\foot{This  follows  from the equivalence of the BES~\cite{Beisert:2006ez} equations  in  the $\mathcal{N}=4$  SYM 
 and  ABJM cases  and  the fact that  $h(\lambda)$  (which is not renormalized in  the SYM case, 
$h_{_{\rm SYM}}={\sqrt{\lambda_{_{\rm SYM}}}\ov 4\pi}$)  but  should be non-trivial   \ci{Gaiotto:2008cg,Nishioka:2008gz} to
correctly  interpolate  between  the weak  and   strong coupling  regimes  in the ABJM magnon dispersion relation   $\epsilon=\frac{1}{2}\sqrt{1+16\, h^2(\lambda)\sin^2\frac{p}{2}}\,.$}
 $f_{0}(\lambda)=\frac{1}{2}\, f_{0_{\rm SYM}}(\lambda_{_{\rm SYM}} )\,$  where one is to  replace ${\frac{\sqrt{\lambda_{_{\rm SYM}} }}{4\pi}\
\rightarrow\  h(\lambda)}\,.$
 According to the conjecture of  \cite{Gromov:2014eha}
 the exact expression for $h(\l)$ is determined by the relation 
 $\lambda = \frac{\sinh{2\pi h(\lambda)}}{2\pi}\, _3F_2 \left(\frac12,\frac12,\frac12 ; 1,\frac32 ; 
-\sinh^2{2\pi h(\lambda)} \right) $, implying that at strong coupling
\be  h(\lambda) ={1\ov \sqrt 2}   \sqrt{ \lambda - \frac{1}{24}} - \frac{\log 2}{2 \pi}  
+ {\cal O}\big( e^{-2\pi\sqrt{2\lambda}} \big), \qquad 
 \lambda \gg 1 \, .\la{91}
\ee
The shift 
$\l \to \l -  \frac{1}{24}$  may be  related to the  redefinition $N \to N - {1\ov 24} ( k - k^{-1}) $
 that  follows  \ci{Bergman:2009zh} from the  presence of the  $R^4\we C_3$  term in the  M-theory effective action. This  shift of $N$ modifies the relation between $\LL$ and $N$
 in \rf{05}  and thus the expressions for $\gs$ and $\T$  in \rf{09}.\foot{Explicitly, we get 
 $ \gs = \frac{\sqrt\pi\, (2\l)^{5/4}}{N}\,\big(1-\frac{1}{24\l}+\frac{\l}{24N^{2}}\big)^{1/4}, \ \ \ 
 \T  = {L^2\ov 8 \pi \alpha'} =   \sqrt{\l\ov 2}\,\big(1-\frac{1}{24\l}+\frac{\l}{24N^{2}}\big)^{1/2}\ $.}
 In particular, one gets
 $ \l= {N\ov k} \to \l - {1\ov 24} ( 1 - k^{-2})= \l - {1\ov 24}  + {\l^{2}\ov 24 N^2}$, 
 where the $1/N^2$   correction   may be ignored  in  the string tree level (planar) approximation. 
 In general, one gets 
 $f(\l) = \sqrt{ 2 \l} + ... \to \sqrt{ 2 \l - {1\ov 12} ( 1 - {1\ov k^2} )} + ... =
  \sqrt{  2 \l - {1\ov 12}}   + {1\ov 12k^2 } \big(    \sqrt {2\l - {1\ov 12} }\, \big)^{-1} + ...
 $.  The resulting corrections  to the coefficients of the $1\ov k^{2n}$   terms 
 in the M2  brane 1-loop contribution in \rf{777} are  thus subleading 
 at   large $\l$   and   will be ignored below.
 They  will    become relevant once the 2-loop $q_1(k)$ term in \rf{0129},\rf{0132}   is taken into account.

\subsection{Non-planar corrections from semiclassical M2 branes  with two spins in $S^7/\ZZ_k$}

One may  apply the same  strategy  of  semiclassically quantizing 
M2  brane solutions  to  find the  leading  strong-coupling asymptotics of  the non-planar corrections to the 
dimensions of other  dual ABJM operators. 
The starting point, like in the familiar 
\adss  case (for a review see, e.g, \ci{Tseytlin:2003ii,
Tseytlin:2010jv}),  is 
a classical   string solution  in AdS$_4 \times \CP^3$  that is dual to a particular 
 ABJM  operator   with large  quantum numbers. 
  One is then to 
 find its   generalization to an M2 brane    in   \adsm  which is  wrapped also 
 on  the 11d circle $\vp$ in  $S^7/\ZZ_k$   \rf{06} (so that upon the 
 double dimensional reduction  or in the $k \to \infty$ limit 
  it reduces to a particular   string solution for which one may   be able 
   to identify the dual  gauge-theory   operator).\foot{Classical rotating  membrane solutions  in flat and AdS  spaces were previously discussed, e.g., in
\ci{Sezgin:2002rt,Axenides:2002wi,Alishahiha:2002sy,Hartnoll:2002th,Bozhilov:2003wr,
Hoppe:2004iu,Brugues:2004pj,Ahn:2008xm,LopezCarballo:2009hog,Axenides:2013lja,Linardopoulos:2015kuh}.}
 
 Considering first the string theory  case, let $\J_r$  be  a collection 
 of parameters (frequencies, etc.) of a classical solution  that are fixed in the large tension expansion. Then   the global  AdS energy   should have the following  large tension expansion 
  \begin{align} 
 \la{92}
 &E=\sqbl\, \E_0 (\J_r) + \E_1(\J_r) + {1\ov \sqbl}\, \E_2(\J_r) + ... \ ,  \qquad \qquad 
 J_r=\sqbl\, \J_r \ ,  \\ 
 &\la{93}\qquad \qquad 
\bl \equiv 2\pi^2 \l \ , \qquad \ \ \ \   \T= {\sqrt{ \l}\ov {\sqrt 2}} = {\sqrt\bl \ov 2 \pi} \ . 
 \end{align}
 To stress the  analogy with the \adss  case
 we  introduced as in \rf{09}  
  the rescaled   coupling $\bl$ (used also in \ci{McLoughlin:2008he}).
In \rf{92} $\E_0$ is the classical contribution, $\E_1$ is  the 1-loop  world-sheet   correction,  etc.
 One can then expand $\E_\ell$  in the  limit  of small or large $\J_r$, 
  express $E$ in terms of  the  spins $J_r=\sqbl \J_r$ and 
  compare to the dimensions of the dual gauge theory operators. The
  1-loop  corrections to energies of  two-spin solutions in  \adsc\   were 
    discussed in  \ci{McLoughlin:2008ms,Alday:2008ut,McLoughlin:2008he,Bandres:2009kw,Beccaria:2012qd}.

  Since here $\sqbl \gg 1$  while $\J_r$ are  fixed,   one has $J_r\gg 1$
   but one may hope that it  may be  possible also   to capture  
     the  strong-coupling behaviour   of dimensions  of   ``short"  operators 
    with finite values of spins (see \ci{Roiban:2009aa,Roiban:2011fe,Beccaria:2012xm,Beccaria:2012qd}).

Starting with  an   M2 brane solution  in AdS$_4 \times S^7/\ZZ_k$ 
that generalizes  a spinning string solution in AdS$_4 \times \CP^3$
the analog of the  large tension expansion in \rf{92}  will be  (cf. \rf{012},\rf{0129})\foot{Here $ {\pi \ov k} \rT_2= \ha  {2\pi\ov k} \T_2$  contains in addition to the factor 
${2\pi\ov k}$  of the   length of the 11d circle in \rf{06} on which the M2  brane is wrapped 
 an extra $\ha$ due to  the  scale of the AdS$_4$  factor in \rf{05}, cf. \rf{011}.}
\begin{align} 
 \la{94}
 &E= {\pi \ov k} \rT_2\, 
  \E_0 (\J_r) + \hE_1(\J_r, k ) +  ( \rT_2)^{-1}  \hE_2(\J_r, k) + ... \ ,  \qquad 
  \qquad  \rT_2 ={k\ov \pi^2} \sqbl \ .
  \end{align}
  In this expansion $\T_2$ is  assumed to be  large 
    while $k$ (the parameter of the 11d background) 
   and $\J_r $ (the parameters of the classical M2 brane solution) are fixed. 
   To relate this  to the small $\gs$  expansion in type IIA string theory in AdS$_4 \times \CP^3$   or  to the large $N$   expansion 
   in the dual ABJM gauge theory we should then  expand $\hE_r$  in large $k$
    for fixed $\J_r$ as in \rf{0132}
    \begin{align} \la{95}
   & \hE_1(\J_r, k ) =\E_1(\J_r) + {1\ov k^2} \cG_{11}(\J_r) + {1\ov k^4} \cG_{12}(\J_r) + ...\ , \\
    & \hE_2(\J_r, k) =
     {k\ov \pi^2}\Big[  \E_2(\J_r)
      + {1\ov k^2} \cG_{21}(\J_r) + {1\ov k^4} \cG_{22}(\J_r) + ...\Big] \ , \ \ \ ... \ . \la{955}
    \end{align}
  The strong-coupling  limit of the  leading non-planar correction  is thus represented by the 
  $\cG_{11}$ term in the 1-loop M2  brane  contribution  $\hE_1$.
  
  One may then   consider the  large or small $\J_r$  limits and 
  finally express the resulting  expressions in terms of the quantum numbers $J_r=\sqbl \J_r$
  to get   predictions for the  corresponding gauge theory anomalous dimensions. 
  The  above order of limits corresponds to operators  for which their quantum numbers  do not grow with $N$, i.e. 
  are fixed in the large $N$ limit  so that $ N^{-1} J_r  \sim {1\ov \sqrt k} \J_r \ll 1$. 
  
    Below  we  will consider the   M2 brane solutions  that  generalize the 
  ``short'' (or  ``slow",  $\J_r \ll 1$)   and  ``long" (or ``fast",  $\J_r \gg 1$)  circular  string solutions with  two equal 
   angular momenta  
   $J_1=J_2\equiv J$  in $\CP^3\subset S^7/\ZZ_k$.
   These   are direct analogs of the  string 
   solutions in \adss for which 1-loop  corrections to energies were discussed  in 
    \ci{Frolov:2003qc,Frolov:2004bh,Beisert:2005mq,Roiban:2009aa}.
    
    The  ``long" $J_1=J_2$   string solution  in AdS$_4 \times \CP^3$  that has  
    the  classical energy $E_0= \sqrt{4 J^2 +  \bl}$
    was already studied  in  \ci{Bandres:2009kw}. Here we will find also 
    its \sho  counterpart with $E_0= \sqrt{2\sqbl J}$.
    The  energies of the   corresponding M2  branes wrapped  on the  11d circle are given by  the same expressions. 
     The   dual  operators  having these  quantum numbers 
        should  be   built out of  the  4  bi-fundamental scalars of the ABJM theory
        as     $\OO= \tr\big[ ( Y^1 Y^\dagger_2)^{J_1} ( Y^3 Y^\dagger_4)^{J_2}\big] +...$.
    
   We will  first  compute  the 1-loop string  corrections  to  the above   classical  energies. 
    In particular,   for the \sho string solution we will find 
    \begin{equation}\label{sm7}
    E_{\text{str}}= 2\sqrt{\sqbl J} 
    +\tfrac{1}{2}+\tfrac{1}{2}\frac{J^{1/2}}{\bar{\lambda}^{1/4}}
    -\tfrac{9}{4}\zeta(3) \frac{J^{3/2}}{\bar{\lambda}^{3/4}}+\mathcal{O}\big(\frac{{J}^{2}}{\bar{\lambda}}\big) \ .
\end{equation}
 This  represents  a prediction for  the  subleading strong-coupling corrections 
    to  the  dimension  $\Delta(J)$    of the  corresponding  dual   ``short" operator
   that has  ``flat-space''  scaling $\Delta \sim \sqrt[4]{\bl}\sqrt J$ \ci{Gubser:2002tv} at  leading order in  strong coupling.
The energy  \rf{sm7}   has a similar structure as the small-spin 
expansion   for the 1-loop corrected   energy of a  short folded  ($S,J)$  string 
spinning  in AdS$_4$ and  also having  orbital momentum  in $\CP^3$  that was found   in 
\ci{Beccaria:2012qd}\foot{For a similar 
expression in the \adss  case see   \ci{Beccaria:2008dq,Tirziu:2008fk,Beccaria:2012qd} (see  also  \ci{Forini:2012bb}).}
\be \la{340}
E_{\text{str}} (S,J)=  \sqrt{2 \sqbl  S} -  \ha   + \tfrac{1}{4}  {({2 S})^{1/2}\ov  \bl^{1/4} } \big[  S^{-1} J (J+1)  + \tfrac{3}{2} S  -1\big]
  + ...\ . 
\ee
It would be interesting to  match \rf{sm7} to  the integrability (quantum spectral curve) 
 strong-coupling predictions for the dimensions of 
the corresponding     states.\foot{For a recent  exposition of the strong-coupling QSC results in the \adss case   see \ci{Julius:2023hre,Ekhammar:2024rfj}.
  }
 These   were  
  found previously  for a few $(S=2,\,  J=1)$  \ci{Gromov:2014eha} 
 and $(S=1,\, J=1,2,4)$ \ci{Bombardelli:2018bqz}  operators   of the form  $\tr \big[ D_+^S   ( Y^1 Y_4^\dagger)^J\big]$ 
  in the $\rm sl_2$ sector of the ABJM theory (see also \ci{Cavaglia:2014exa,Bombardelli:2017vhk}).\foot{The BMN  operator   which is  the vacuum in the $\rm sl_2$ sector 
    corresponds to $\tr    ( Y^1 Y_4^\dagger)^J$  in the representation $[J,0,J]$ of $SU(4)$. For  early  discussions  of integrability of the  ABJM theory see   \ci{Minahan:2008hf,Gaiotto:2008cg,Klose:2010ki}.}

  We will then generalize the string 1-loop computations  to the M2 brane  ones  
  getting predictions for the  non-planar corrections to the dimensions of the above $J_1=J_2$  
   operators  at strong coupling.
    For the ``short"  M2 brane solution   we will find the following $1/k^2$ correction to \rf{sm7} 
  \begin{align}
\label{sm7778}
    E_{_{\rm M2} }= &2\sqrt{\sqbl J} 
    +\tfrac{1}{2}+\tfrac{1}{2} \bar{\lambda}^{-1/4} J^{1/2}
    -\tfrac{9}{4}\zeta(3) \bar{\lambda}^{- 3/4}  J^{3/2}+\mathcal{O}\big(\bl^{-1} {J}^{2}\big) \no\\
 & +   {1\ov k^2} \Big[ \zeta(2)  \big( -4 \bl^{3/4} J^{-3/2}  +8   \bl^{1/4} J^{- 1/2}  \big) + \OO\big(\bl^{-1/4} J^{1/2} \big)\Big]  + \OO( {1\ov k^4}) 
    \ .  \end{align}
  From the string theory point of view   the 
   membrane correction term   $\sim {1\ov k^2} = {\gs^2\ov 4 \sqbl} $   
    represents  the leading  large tension  asymptotics 
   of the  string 1-loop (torus)  contribution.  
   
    On      the dual ABJM gauge theory side \rf{sm7778}
    should be  understood as the expansion  first in $1/N^2$  and then in large $\l$ for fixed
 quantum number   $J$. The 
         ${1\ov k^2} = { \l^2\ov N^2}= {\bl^2\ov (2\pi^2)^2 N^2}$ 
     term in \rf{sm7778}  then  represents  a prediction 
   for the    leading non-planar correction 
  to  the  dimension  of the corresponding   \sho operator. 
 
In  the \lon M2 brane case we will find 
\begin{align}
    E_{_{\rm M2}}= 2J  &+  
     \tfrac{1}{4} \bl J^{-1} (1 - 2 \log 2\,  \bl^{-1/2} + ...)  + \ha  c_1  \bl  J^{-2}
      ( 1 + ...)  + ...\no \\
    &+  {1\ov k^2} \,  \zeta(2)\big{(}-8\bl^{-1/2}  {J} - 2  {\bl^{1/2}  }J^{-1} +\tfrac{3}{16} \bl^{3/2} {J}^{-3} + ... \big{)}
    + \OO({1\ov k^4}) \ . \la{381} \end{align} 
      Here  $c_1\approx -0.336$   and the ${1\ov k^2} = {\bl^2 \ov (2 \pi^2)^2 N^2}$  term represents   a
       prediction for the  strong-coupling limit of the  leading 
    non-planar correction   to the dimension of  the corresponding operator  with the large  spin $J$. 
    
\

 The rest of this   paper is organized as follows.
In   section 2 we will present   the  1-loop  M2   brane  computation 
 that generalizes the  leading   strong coupling \adsc\ string theory  contribution to the  ABJM cusp anomaly 
   to the non-planar level. 

 In section 3 we will consider the  M2 brane  generalizations
  of the ``long" and ``short"  circular string solutions
 with equal  spins in  $\CP^3$  and compute the corresponding 1-loop  corrections 
 to the AdS$_4$ energies in the large $k$ expansion. 
 
 Some  open problems will  be mentioned  in section 4.
 There  are also several Appendices  containing some  details of  the computations.   
 In Appendix \ref{ap4}  we will make  some comments on  non-planar corrections
 to  the multi-wound Wilson loop and the \brf in the ABJM  theory related to the discussion in section 4.

\renewcommand{\theequation}{2.\arabic{equation}}
 \setcounter{equation}{0}

\section{
 1-loop correction  to energy of   M2 brane spinning in AdS$_4$}

In this section  we will  compute   the  1-loop  correction to the 
 partition function  \rf{036} 
 expanded near  the classical M2    brane solution in AdS$_4 \times S^7/\ZZ_k$ 
 that generalizes the 
infinitely long rotating folded string  \ci{Gubser:2002tv,Frolov:2002av,McLoughlin:2008ms}   in  AdS$_4$.
This will determine  the function $q_0(k)$ in \rf{0129}-\rf{777}, i.e. 
the   leading  large $\lambda$ corrections  at each order in the $1/N^2$  expansion 
in  the cusp  anomalous  dimension    in the ABJM theory.

In terms of the AdS$_4 \times S^7/\ZZ_k$  coordinates in \rf{100s},\rf{06} the  relevant  large-spin  (infinitely long) 
membrane   solution is 
\begin{align}
t=\ka\, \s^0, \ \qquad \r= \ka\, \s^1, \ \qquad \a=0\ , \qquad  \b= \ka\, \s^0, \ \qquad   \vp= \s^2  \ ,   \la{2s}
\end{align}
with $\CP^3$  coordinates in \rf{07}  being trivial  
and $\ka$   being a  constant parameter.
 Here $\xi^i$ ($i=0,1,2$) are the membrane world-volume coordinates  with  $\s^2\in (0, 2\pi)$.
 One of 4 segments of the folded  closed string  is represented  by  $\s^1 =(0,{\pi\ov 2})$.
 We will consider the  limit $\k\to \infty$   in which the rescaled $\s^1 \to \k \s^1$   can be decompactified. 
  The corresponding  classical   AdS$_4$ energy and the  spin satisfy  (${\cal S}  ={S\ov  \sqbl } \gg 1$) 
\be 
E_0 - S = {1\ov 4} {(2\pi)^2 \ov  k} \T_2 \,   \ka = \sqbl\, \ka= \sqrt{2\l}\, \log S    \ , \ \ \ \ \ \ \ \ \ \ \  \ka= {1\ov \pi} \log S\gg 1 
\ . \la{2.1}
\ee
The dependence on the parameter $\ka$  can be scaled away 
 by  redefining  the coordinates 
$\s^0,\s^1$;  it will  then appear only as  an 
overall factor in the  log  of the quantum partition function or  in the quantum correction to the 
 energy.  It is useful also to perform the  Euclidean continuation $\s^0\to i \s^0$.
  The resulting  induced 3d metric is flat    (cf. \rf{05})
\be 
g_{ij} =\four { \LL^2 } \bar g_{ij}, \ \  \qquad   \ \ \   \bar g_{ij}=   ( 1, 1, {4\ov k^2}) \la{300}\ . 
\ee 
 We will expand  all 3d  fluctuation fields in Fourier modes in $\s^2=\vp$  
  getting an effective 2d field theory on 
  $\mathbb R^2$  with $l=0$   sector representing the modes
   of the type IIA string on AdS$_4 \times \CP^3$  and the  $l\not=0$  tower   being the  genuine 
    membrane  modes.
  The  derivation of the  corresponding 
    fluctuation operators in \rf{037}   is  very similar to the  case of the 
  AdS$_2 \times S^1$   M2 brane representing  the circular Wilson loop 
  that was  discussed in  \ci{Sakaguchi:2010dg,Giombi:2023vzu}.

We will  fix the static gauge  setting to zero the fluctuations of $t,\, \r$ and $\vp$ 
  so that  the non-zero bosonic  fluctuations will be those of 
$\td \a=  \a$,  $ \td \b =\b - \s^0$  and of the  6 real  $\CP^3$   coordinates.
Extracting the overall  factor  $\four { \LL^2 } $  the  resulting  fluctuation   operators will contain 
the ``free"  part 
\be 
- \bar g^{ij} \del_i \del_j =- ( \del_0^2 + \del_1^2 + \four  k^2  \del_2^2 ) \ \  \ \to \ \ \ \ 
 p^2  + \four  k^2 \nn^2 \ , \ \ \ \ \ \  p^2=p_0^2 + p_1^2\ , \ \ \ \ \   \nn=0, \pm 1, \pm 2, ... \ ,  \la{40}
\ee
plus effective mass terms.  
Here  $p_\a$ are   the  momenta
 in the non-compact $\s^0$ and $\s^1$  directions 
  and $l$ is the mode number in the circular $\s^2$ direction.  
One  finds that the 6 real $\CP^3$   fluctuations  have masses
\begin{equation}
m_{\nn}^2 = \four k\nn(k\nn+2)\, ~~~~~~~~(\mbox{6 modes}),\la{5}
\end{equation}
 where  the  linear in $ k \nn $ term comes from the mixing   between the   constant 
 $d \vp $  term and  
 $ k \rm A$  (which is quadratic in fluctuations)  in the 
 $S^7/\ZZ_k$   metric in  \rf{06}.
 The  $\nn=0$  modes in \rf{5}   are 6  massless  excitations 
  in the  corresponding folded string spectrum  in AdS$_4 \times \rm CP^3$.
 Note that  if  $k=1$  we  get  6 
 tachyonic modes with $\nn=-1$ indicating an  instability of the membrane 
 wrapped on a  big circle  of $S^7$.\foot{The  classical spinning membrane solution in AdS$_4\times S^7$  
     that corresponds to   a folded spinning  string in AdS$_4$  
was discussed also  in \ci{Axenides:2013lja}  and   this reference had comments on 
its instability by analogy with a string  wrapped on a circle in the sphere.}
As we are interested in the large $k={N\ov \l}$ expansion, below we  will 
 assume that $k>1$  but will   return to  the $k=1$  case 
 at the end of this section. 

 Expanding  the volume $\int \sqrt{g}$   part \rf{013} of the membrane action 
 we get for the quadratic  Lagrangian for the 2 remaining bosonic  3d  fluctuations
  $ \td \a$ and  $ \td \b$  (scaling  out  the $\four\LL^2$ factor  in \rf{300}) 
 \be \la{8}
 L_{\rm V}= \ha  \Big[   \sinh^2 \r\,  \cosh^2 \r\,   \del^i \td \b   \del_i \td \b      +    \sinh^2 \r\,  ( \del^i \td \a  \del_i \td \a    +  \td \a^2) \Big] \ , 
 \ee
 where $\r=\s^1$.  After the 3d  field redefinition $(\td \b , \td \a) \to (u,v)$
 \be \la{9}
 \td \b = (  \sinh \r\,  \cosh \r)^{-1} u  \ , \qquad  \ \ \ \ \ \  \td \a =   (  \sinh \r )^{-1} v   \ , 
 \ee
 and integrating  by parts 
    we get (as in the   static-gauge analysis  in the  \adss  case \ci{Frolov:2002av}) 
 \be \la{1001} 
   L_{\rm V}= \ha  \big(   \del^i u    \del_i u     + 4 u^2    +    \del^i v \del_i v   +  2 v^2  \big)\ . 
   \ee
   In addition, there is a  contribution coming from the WZ  term in the membrane action \rf{032}  with $C_3$ given by \rf{7}.  Using \rf{2s}  it  leads to 
 the  mixing term $ L_{\rm WZ} \sim v \del_2 u$  
 so that  in total we get 
 \be \la{10} 
  L(u,v)= L_{\rm V} +  L_{\rm WZ}  = \ha  \big(   \del^i u    \del_i u     + 4 u^2    + 
   \del^i v \del_i v   +  2 v^2  \big)  - 3 v \del_2 u \ . 
   \ee
   Expanding in modes  in $\s^2$  we  have 
 $\del_2 \to \ha  i {k \nn}$   and thus  diagonalizing \rf{10}   find  two  towers of 2d scalars   with the following  masses (which are positive for any $k l$) 
 \be 
 \la{11} 
\te  m^2_{\nn,+} =  3  + {1\ov 4}  k^2 \nn^2  +   \sqrt{ 1 +  {9\ov 4}  k^2 \nn^2  } \ , \qquad \qquad 
 m^2_{\nn,-} =  3  + {1\ov 4}  k^2\nn^2  -   \sqrt{ 1 +  {9\ov 4}  k^2 \nn^2  }\ .
 \ee
  For $\nn=0$ we reproduce  the values of  masses (4 and 2) of the two AdS$_4$  fluctuations 
  in the  corresponding string theory  limit.\foot{Note that the mass of  the fluctuation of the coordinate  $\a$  transverse to the AdS$_3$  subspace where the string moves is the same 2
    as in the  case of the AdS$_5 $ string solution (where there are two such modes).
 The only mass that changes   is that of  the  fluctuation of $\b$
 as  the string  is  rotating in this direction. In general, the mass of such mode is 
 $4 + R^{(2)}$   where $R^{(2)}$ is the curvature of induced metric.
  For comparison, in the case 
  of the  AdS$_2 \times S^1$  membrane in    \ci{Giombi:2023vzu} 
  the   mass terms   of the corresponding   fluctuations  in \rf{10} were both   equal to 2 
  (due to  the shift of  mass term 4 by the  scalar curvature $R^{(2)}=-2 $ of AdS$_3$)
  and then  $m^2_{\nn,\pm}  =  2  + {1\ov 4}  k^2 \nn^2  \pm     {3\ov 3}  k  \nn $. 
  As $\nn$   takes both positive and negative values this is  equivalent to  having 
   2 modes with $m^2_\nn = {1\ov 4} (k\nn-2) (k\nn -4)$. 
  }

 Finding the quadratic fermionic Lagrangian  from \rf{33}  is very similar  to   the 
 AdS$_2 \times S^1$ membrane case   \ci{Giombi:2023vzu}  and one  gets 
  8  fermionic towers  in flat 2d space with masses
 \begin{equation}
m_\nn  = \ha {k\nn}\pm 1 ~~~(\mbox{3+3 modes})\,,\qquad \qquad
m_\nn =\ha {k\nn} ~~~(\mbox{2 modes})\ . 
\label{6}
\end{equation}
For $l=0$   this  reproduces the spectrum of the  fermionic fluctuations 
for the infinite  folded  string in AdS$_4 \times \rm CP^3$ \ci{McLoughlin:2008ms,Alday:2008ut}.

Integrating out   8+8  towers of  fluctuations in $\mathbb R^2$ 
 and summing over $l$   we then get  the 1-loop   partition  function $\Z_1$  in \rf{037}.
 Since all fluctuation operators  have constant coefficients,   
 $\log \Z_1$    will be proportional to the  3d volume  containing the  $\ka^2$  factor  from rescaling of 
 $\s^0$ and $\s^1$ (cf. \rf{2s}).  The 1-loop  correction to  the 
 world-volume energy   will then scale as $\ka^2$. Since $t=\ka \s^0$  the  corresponding correction to 
 the  AdS$_4$  energy will scale as 
 $\ka= {1\ov \pi} \log S$ leading to the  expression for the 1-loop term $q_0$ in the scaling function  
 $f(k,\T_2)$ in \rf{0129}. Explicitly, we find 
 \begin{align}
 \Gamma_1= &-\log \Z_1 = \ha  V q_0  \ , \ \ \   \ 
   \qquad  V=  \ka^2 \int d\s^0d\s^ 1 \ , \qquad \qquad E_1= \pi q_0 \ka= q_0 \log S  \ ,  \la{15} \\
   q_0  =&  
   \int  {d^2 p \ov (2 \pi)^2 }  \ \Big[  Q_0(p^2)  + 2 \sum_{\nn=1}^\infty  Q_\nn(p^2)\Big]
   = p_{00} + \bar q_0(k)  \ , \la{1555}
   \\
   Q_\nn(p^2)=& \textstyle  \log\big[ p^2 + 3  + {1\ov 4}  k^2 \nn^2  +   \sqrt{ 1 +  {9\ov 4}  k^2 \nn^2  }\big]
 + \log\big[ p^2 + 3  + {1\ov 4}  k^2 \nn^2  -   \sqrt{ 1 +  {9\ov 4}  k^2 \nn^2  }\big]\no \\ 
 & \textstyle  + 3  \log\big[ p^2 +  \frac{1}{4}(k\nn)^2 + \frac{1}{2} k\nn  \big] 
  + 3  \log\big[ p^2 +  \frac{1}{4}(k\nn)^2 - \frac{1}{2} k\nn  \big] \no \\
 &\textstyle  - 3    \log\big[ p^2 + (1 +  \frac{1}{2}  k\nn )^2\big]  -  3    \log\big[ p^2 + (1 -  \frac{1}{2}  k\nn )^2\big]
  - 2   \log\big[ p^2 + (  \frac{1}{2}  k\nn )^2\big] \ . \la{16}
 \end{align}
 Here  in \rf{1555}  we  followed \rf{045}  and separated  the $k$-independent   contribution $p_{00}$ 
 to $q_0$   coming from the $l=0$ (string-theory)  part $Q_0$  of the integrand.
 Computing the 2d momentum integral  
 gives 
 \be\la{18}
p_{00}=  \int { d^2p \ov (2\pi)^2}  \, Q_0 (p^2) =  \int^\infty_0  {d p^2\ov 4 \pi}  \ \Big[ 
  \log ( p^2 + 4)  + \log(p^2+2 )    + 4  \log  p^2       - 6    \log( p^2+1) \Big] = - {5\log 2\ov 2 \pi}  \ , 
 \ee
 thus reproducing the value of the 1-loop  correction to the cusp  anomaly 
 in string theory in AdS$_4\times \CP^3$  given in \rf{555}. 
 
 The integral of   $Q_\nn$  with $\nn>0$  giving $\bar q_0(k)$ in \rf{1555}   is also   UV finite 
 (as one   can check  explicitly  by   doing  the integral  over $p^2$  between 0 and $\Lambda$
  and  taking the limit $\Lambda \to \infty$) 
 \begin{align} 
&\qquad \qquad \qquad  \bar Q_\nn \equiv  \int^\infty_0 { d p^2\ov 2 \pi} \,  Q_\nn(p^2) \no \\
&= 
 -\frac{1}{8\pi} \Big[-  3 (k\nn-2)(k\nn-4)   \log ( k\nn -2)  -   3 (k\nn+2)(k\nn+4)   \log ( k\nn +2) 
   + (k\nn)^2  \log [(k\nn)^2] \no  \\ 
 &  + \big[(k\nn)^2 + 12 \big]  \log\Big[ (k\nn)^4 - 12 (k\nn)^2 + 128 \Big] + 2 \sqrt{9 (k\nn)^2+4} \log { (k\nn)^2+12 + 2 \sqrt{9 (k\nn)^2+4} \ov  (k\nn)^2+12 - 2 \sqrt{9 (k\nn)^2+4} }\
\Big]\ . \la{20}
  \end{align}
 Expanding  in    large $k$ then gives 
 \be 
 \bar Q_\nn = \frac{4}{\pi  (k\nn)^2}  +  \frac{4}{\pi  (k\nn)^4}-\frac{1616 }{15 \pi  (k\nn)^6}-\frac{38944}{35 \pi  (k\nn)^8}
-\frac{447488}{105 \pi  (k\nn)^{10}}+\frac{2227200}{77 \pi  (k\nn)^{12}}+ ... \ . 
\la{22}
 \ee
 The remaining sum over M2  modes  in \rf{1555}  thus    converges, 
  leading to  \be 
 \bar q_0 = \sum_{l=1}^\infty \bar Q_\nn  
  =\frac{2 \pi }{3 k^2} +\frac{2 \pi ^3}{45 k^4 } -\frac{1616 \pi ^5}{14175 k^6}
 -\frac{19472 \pi ^7}{165375 k^8} 
 -\frac{447488 \pi ^9}{9823275 k^{10}}  + \frac{20519936 \pi ^{11}}{655539885 k^{12}} + ...
 \la{211}\ . 
  \ee
  This determines  the non-planar  coefficients $p_{0r}$ in $\bar q_0$ in  \rf{045}
  in terms of the  values of $\zeta(2m)= \sum^\infty_{\nn=1} {1\ov \nn^{2m}}$,  thus 
  reproducing  \rf{777}. 

 In the above    derivation of \rf{20} we assumed that $k  >1$  when 
  \rf{20} 
  is real.  
   Analytically continuing  $\bar Q_\nn$ to 
       $k=1$ we get an imaginary part\footnote{The imaginary part is equal to $9/8$ and arises from the $\nn=1$ term. The real part is obtained by evaluating the sum numerically.}  
 \be\la{222}  
 \ \bar q_0\big|_{k=1} = -0.663 + 1.125 i \ . \ee
As already noted   below eq.\rf{5}, 
this reflects an instability of the 
membrane  that rotates only in AdS$_4$  and 
is   wrapped on a circle inside   $S^7$  
which is contractable. 


\renewcommand{\theequation}{3.\arabic{equation}}
 \setcounter{equation}{0}
\section{
  M2 branes rotating   in $S^7/\ZZ_k$ }

Let us  now  provide an illustration of the strategy described in section 1.4
and consider  1-loop corrections  to  the two membrane  solutions  that generalize 
the \sho and \lon  circular string solutions  with two angular momenta $J_1=J_2$ 
in $\mathbb R_t \times \CP^3$  part of \adsm.\foot{These are direct counterparts 
of the  string solutions  in \adss  describing  a  rigid   circular  string rotating in two orthogonal planes in $S^5$  with $J_1=J_2=\sql \J$ having two   branches:
 \lon  one with $\J \geq \ha$ and \sho  one  with $\J \leq \ha$
 \ci{Frolov:2003tu} (see also \ci{Frolov:2003qc,Arutyunov:2003za}).
While  the radius of the  \lon  string    is fixed to be that of  $S^5$  so it is never small  and 
 admits a ``fast-string'' expansion  $\J = { J\ov \sql} \gg 1$, the \sho one 
  may  have an arbitrarily  small  radius and spin  and thus has 
 a   ``slow-string'' limit  $\J \ll 1$  when 
    it probes the near-flat region   of $S^5$. 
These are  among the  simplest rigid string solutions   with  
 explicitly known   spectrum of small fluctuations. For the \lon  branch   the  1-loop corrections to the  energy were
  computed in the large $\J$  expansion 
  \ci{Frolov:2003qc,Frolov:2004bh,Beisert:2005mq},   with ``non-analytic"  terms found in \ci{Beisert:2005cw,Minahan:2005qj}.
  The 1-loop correction to the energy of   the \sho   solution was  found 
   in \ci{Roiban:2009aa,Roiban:2011fe,Beccaria:2012xm}. Note that  in addition  to the circular  solution 
    there is  also a
       folded string  solution   with 2 spins in $S^5$ 
  that has less energy  for given values of spins  \ci{Frolov:2003xy}.
   The  study of these  simplest   solutions played an important role in establishing the integrability approach to the spectrum of strings in \adss.
  }
 We shall first   describe these     string solutions in \adsc\ 
 (with the \lon one  previously found in \ci{Bandres:2009kw})
 and then generalize them to M2 brane  solutions in \adsm.
    The M2 branes will be located at the center  of AdS$_4$ with $t=\xi^0$, wrapped on 
  the  11d circle $\vp$ in \rf{06}   and rotating in $\CP^3$. 
  We shall then  compute  the  1-loop corrections to the energies of these \sho and \lon M2 brane solutions  and  study their 
  expansions  in  spins   and  11d radius ${1\ov k}$.
    
\subsection{Classical solutions}

It will be useful  to use the explicit  parametrization of $S^7/\ZZ_k$  in terms  of the 7 angles   choosing   
the 4 complex coordinates  subject to $z_a\bz_a=1$  in  as 
\begin{align}\label{A1}
    &\te z_{1}=\cos\chi\cos{\theta_1\ov 2}\, \exp\big[ i(\frac{\y}{k}+\frac{\psi+\phi_{1}}{2})\big] \ , \qquad \qquad z_{2}=\cos\chi\sin{\theta_1\ov 2}\, \exp\big[ i(\frac{\y}{k}+\frac{\psi-\phi_{1}}{2})\big] \ , \nonumber \\ 
    &\te z_{3}=\sin\chi\cos{\theta_2\ov 2}\, \exp\big[i(\frac{\y}{k}-\frac{\psi-\phi_{2}}{2})\big] \ , \qquad \qquad z_{4}=\sin\chi\sin{\theta_2\ov 2}\, \exp\big[i(\frac{\y}{k}-\frac{\psi+\phi_{2}}{2})\big] \ , 
\end{align}
so that the $S^7/\ZZ_k$   metric in \rf{07}  takes the form 
\begin{align}\label{A12}
   &ds^2_{S^7/\ZZ_k }=ds^2_{{\CP}^{3}}+\frac{1}{k^2}(d\y+k\AA)^2 \ , \ \ \ \qquad \te \AA=\frac{1}{2}\big[\cos(2\chi)d\psi+\sin^2\chi\cos\theta_2d\phi_{2}+\cos^2\chi\cos\theta_1d\phi_{1}\big] \ , \nonumber \\
    &ds^2_{{\CP}^{3}} =\te  d\chi^2 + \cos^2\chi\sin^2\chi\big{(}d\psi+\frac{1}{2}\cos\theta_1d\phi_{1}-\frac{1}{2}\cos\theta_2d\phi_{2}\big{)}^2\\ \nonumber
    & \qquad \qquad \qquad  \qquad \qquad+\te \frac{1}{4}\cos^2\chi\big{(}d\theta_{1}^2+\sin^2\theta_1d\phi_{1}^2\big{)}+\frac{1}{4}\sin^2\chi\big{(}d\theta_{2}^2+\sin^{2}\theta_2d\phi_{2}^2\big{)} \ .
\end{align}
Here $\chi\in[0, \pi/2), \ \y \in [0, 2\pi), \ \psi \in [0, 2\pi), \ \theta_{i}\in[0, \pi), \ \phi_{i}\in [0, 2\pi)$.

\subsubsection{String solutions}

Starting with the bosonic part of the   string   action in \adsc\ we shall fix  the conformal gauge.
Then   the relevant $\mathbb{R}_t\times\CP^3$ part of the  action  may be written 
in terms of that of the $\CP^3$ sigma model as 
(cf. \rf{07},\rf{09})
\begin{equation} \label{s2}
    S_{\rm str} =- 2 \T  \int d^2\xi \ \Big[-\four (\partial_{\alpha}t)^2 + |D_{\alpha}z^{a}|^2 - \Lambda(\xi)\big(|z^{a}|^2-1\big)\Big]  \ , \qquad 
    \qquad \T= {\sqbl \ov 2 \pi} \ . 
\end{equation}
Here $\alpha =(0, 1)$, $\Lambda(\xi)$ is a Lagrange multiplier imposing the  $\bz_a z_a=1$   constraint on 4 complex coordinates $z_a$. $D_{\alpha} $  is a  $U(1)$  covariant derivative 
containing  an auxiliary gauge field $A_\a$  
\begin{equation}\la{3.2} 
    D_{\alpha}z^{a} = \partial_{\alpha}z^{a}-iA_{\alpha}z^{a} \ , \ \ \ \ \ \ \ \ \ \ \ 
    z^{a} \rightarrow e^{i \eps}z^{a} \ , \ \ \ \ \ \ \ A_{\alpha} \rightarrow A_{\alpha} + \partial_{\alpha}\eps \ , \ \ \ \ \ \   \eps=\eps(\xi) \ .
\end{equation}
  The equations of motion  that follow  from \rf{s2} are
\begin{align}\label{s3}
&        \partial_{\alpha}\partial^{\alpha}t = 0 \ ,  \qquad  \qquad 
                D_{\alpha}D^{\alpha}z^{a} =- \Lambda z^{a} \ , \qquad 
            \ \                 \Lambda =   |D_{\alpha}z^{a}|^2  \ , \qquad 
 \qquad 
         \bar{z}_{a}z_{a} = 1 \ ,  
\\ \label{s30}
  &  A_{\alpha} = \tfrac{1}{2i}\big{(}\bar{z}_{a}\partial_{\alpha}z^{a} - z^{a}\partial_{\alpha}\bar{z}_{a}\big{)} \ ,
\qquad \qquad    |D_{\alpha}z^{a}|^2 = \eta^{\alpha \beta}\big{[} \partial_{\alpha}\bar{z}_{a}\partial_{\beta}z^{a}-(\bar{z}_{a}\partial_{\alpha}z^{a})(z^{b}\partial_{\beta}\bar{z}_{b})\big{]} \ .
\end{align}
We thus   get the  expressions  that correspond to the metric  \rf{07}  (with  $ A_\a $ related  to  the 1-form $\AA$ 
 and  $z^a$  being the embedding coordinates of $S^7$). 
In addition, we have the 
conformal gauge  constraints ($g_{\alpha \beta}$ is the induced metric)
\begin{equation}\label{s70}
  g_{00} + g_{11} =0 \ , \ \ \ \   \ \  g_{01}=0 \ , \ \ \ \ \ \ \ \ 
    g_{\alpha \beta} = -\tfrac{1}{4}\partial_{\alpha}t \partial_{\beta }t + (D_{(\alpha}z^{a})^{\dagger}D_{\beta)}z^{a} \ .
\end{equation}
The action \eqref{s2} is invariant under the global $SU(4)$ symmetry. We 
may choose its  Cartan  generators  as 
\begin{equation}
    H_1 = \tfrac{i}{2}\, \text{diag}(1, -1, 0, 0) \ , \ \qquad  \ H_{2} = \tfrac{i}{2}\, \text{diag}(0, 0, 1, -1) \ , \ \qquad  \ H_{3} = \tfrac{i}{2}\,\text{diag}(1, 1, -1, -1) \ , \la{3.8}
\end{equation}
which    correspond to the Killing vector fields $\partial_{\phi_1}, \  \partial_{\phi_2}$ and $\partial_{\psi}$  of \rf{A12} respectively. 
The associated conserved charges or angular momenta
   and the AdS$_4$ energy then are 
   \begin{equation} \label{s6}
    J_r =  2\T  \int_{0}^{2\pi} d \xi_1 \ \Big{[} (D_{0}z)^{\dagger}H_r z - z^{\dagger}H_r D_{0}z\Big{]} \ , \qquad \qquad\qquad    E_0 = \T \int_{0}^{2\pi}d\xi_{1} \ \partial_{0}t \ .
\end{equation}
We shall  consider  a   class of ``rigid" string  solutions  for  which (cf. \rf{A1}; $a=1,...,4$)
\begin{equation}\la{310}
    t = \kappa\, \xi^{0} \ , \ \qquad \qquad  \ z_{a}=r_{a}\ e^{i\g_a (\xi)}  \ , \ \ \ \ \ \ \
    \ \ \ \ 
    \g_a= w_{a}\xi^{0}+m_{a}\xi^{1} \ , 
\end{equation}
where $r_a, \ w_a$ and $m_a$   are  constant  ``radii",   frequencies and winding numbers. 
Fixing the $U(1)$ gauge symmetry of \rf{s2}  by the  $A_0=0$ condition,\foot{Note that under the gauge transformation \rf{3.2}  the phases $\g_a(\xi)$ in \rf{310}  are  all shifted   by $\eps(\xi)$. 
Thus a  solution found in the $A_0=0$ gauge  that may have $\sum_{a}\g_{a}\not=0$ may be transformed
 into a gauge   where $\sum_{a}\g_{a}=0$. Note also that the charges \rf{s6} are invariant under the gauge transformation \rf{3.2}.} 
the equations of motion \eqref{s3} together with the Virasoro constraints \eqref{s70} reduce to the system of  algebraic equations 
 on the parameters in \rf{310} 
\begin{align}\label{s8}
    &  A_0=0 , \ \ \ \ \ \ \ \ \ \ A_{1}=\sum_{a}r_{a}^2m_{a} \ , \qquad \ \ 
          -w_{a}^2+(m_{a}-A_{1})^2 = \Lambda  \ , \qquad \ \ 
        \sum_{a}r_{a}^2w_{a}=0 \ ,   \\
       &\la{312}
         \qquad 
        \sum_{a}w_{a}(m_{a}-A_{1})r_{a}^2 = 0 \ ,  \qquad \qquad \ \ 
        \tfrac{1}{4}\ka^2=\sum_{a}r_{a}^2(m_{a}-A_{1})^2+\sum_{a}r_{a}^2w_{a}^2 \ , \qquad \ \ \   \sum_{a}r_{a}^2=1 \ .
\end{align}
Evaluated on \rf{310}  the angular momenta in \rf{s6}  may be written as 
\be\la{313}
    J_{1}=4\pi \T \big{(}w_{1}r_{1}^2-w_{2}r_{2}^2\big{)} \ ,  \qquad  \ J_{2}= 4\pi \T\big{(}w_{3}r_{3}^2-w_{4}r_{4}^2\big{)} \ , \qquad 
    J_{3}=4\pi \T \big{(}w_{1}r_{1}^2+w_{2}r_{2}^2-w_{3}r_{3}^2-w_{4}r_{4}^2 \big{)} \ .
\ee
We  shall consider two special  solutions of \rf{310}--\rf{312} for which 
$J_{1}=J_{2}$,   $J_{3}=0$. 

The first  is the \sho one  
\begin{align}
&r_1=r_2\equiv a \ , \qquad r_3=r_4= \sqrt{\ha - a^2} \ , \qquad    m_1=m_2=-m_3=-m_4 \equiv \ha m  \ ,  \la{314}\\
& w_1=-w_2= 2m ( \ha - a^2)\ , \qquad w_3=-w_4 = 2 m a^2 \ ,  \qquad 
\ka^2 = 32m^2a^2(\tfrac{1}{2}-a^2)\ ,\no \\
& A_0=0, \ \ \ \ \ \ A_1= 2m(a^2-\tfrac{1}{4})\ ,  \qquad  \ \ \  g_{\a\b}=  \bc^2 \eta_{\a\b} \ , \qquad  \bc^2= \tfrac{1}{8} \ka^2 \ , \qquad \ \  \Lambda=0 \ . \la{3144}
\end{align}
 Explicitly,  for $z^a$ in \rf{310} we get 
 \begin{align}
&  z^{1}=a\ e^{i[m ( 1 - 2a^2)\xi^{0}+\frac{1}{2}m\xi^{1}]} \ , \ \qquad  \qquad\ \ \ \   \ z^{2}=a\,  e^{i[- m ( 1 - 2a^2)\xi^{0}+\frac{1}{2}m\xi^{1}]} \ ,\no \\
& z^{3}=\sqrt{\ha -a^2} \  e^{i[2m a^2\xi^{0}-\frac{1}{2}m\xi^{1}]} \ , \ \qquad \qquad  \ z^{4}=\sqrt{\ha -a^2} \ e^{i[-2m a^2\xi^{0}-\frac{1}{2}m\xi^{1}]} \ , \la{315}
 \end{align}
or, equivalently, in terms of the  $\CP^3$ angles in \rf{A1},\rf{A12}
\begin{equation}
    \cos \chi_{0} =\sqrt{2}\, a \ ,\ \ \  \ \ \ \theta_{1}=\theta_{2}=\tfrac{\pi}{2} \ , \ \ \  \ \ \ \psi = m\xi^1 \ ,\ \ \ 
    \ \ \  \phi_{1}=4m ( \ha - a^2) \xi^{0} \ , \ \ \ \phi_{2}=4 m a^2 \xi^{0} \ . \la{316}
\end{equation}
 Here $0\leq a\leq\frac{1}{\sqrt{2}}$ 
 and $m$ is the  winding number that  takes integer values.\foot{As $\xi^1\in (0, 2\pi)$   one could  think that $m$  should take only even values. Note, however,  that under  
$\xi^1 \rightarrow \xi^1+2\pi$  we get  $z^{a} \rightarrow e^{\pm i m\pi} z^{a}$
which for any integer $m$  is just an overall phase   of $z^a$   which is a trivial   symmetry of  $\CP^3$ (global part 
of the $U(1)$ gauge symmetry).}   
The corresponding charges are 
\be \la{s9}
J_3=0 \ , \ \ \    J_{1}=J_{2} \equiv J=\sqbl\, \mathcal{J}  \ , \qquad  \J =8 m a^2(\tfrac{1}{2}-a^2)= \four m^{-1}  \ka^2 \ ,  \ \qquad 
  \ E_0 = \sqbl\, \kappa = \sqrt{4m\sqbl\, J}\ .
\ee
Here the  spin    is bounded, i.e. $0\leq \J\leq \ha m  $ or 
    $0\leq J\leq \ha m \sqbl $, with  the maximum at $a=\ha$ and the minimum at 
$a=0$ or $a={1\ov \sqrt 2}$.
Note   that like  for the analogous   solution in  \adss     \ci{Frolov:2003qc}
 the relation between the   energy  and  spin    is the same as   for  the corresponding  solution in flat space 
(i.e.  for a circular string  rotating in 2 orthogonal planes in $\mathbb R^4$). 

 To see  that for  $a\to 0$   the solution reduces to its flat-space analog 
one is to do a $U(1)$ gauge transformation $z_{a}\rightarrow e^{i\frac{m}{2}\xi^{1}}z_{a}$ 
 that   sets  the $a\to 0$ value of $A_1$ (equal to $-\ha m$ in \rf{314}) 
to  zero.\foot{\la{f1}Note that  since $\xi^1 \in [0, 2\pi)$  one cannot in general  transform a constant $A_1$ component of the 
 potential in \rf{s8},\rf{314}    to zero  if ${1\ov 2\pi}\int d \xi^1\, A_1$ is not a  (half) integer.
 }
Then 
\begin{align}\la{319}
a\to 0: \qquad \qquad 
    z^{1}\rightarrow a \, e^{im(\xi^{0}+\xi^{1})} \ , \qquad \  z^{2}\rightarrow a \, e^{im(-\xi^{0}+\xi^{1})} \ , \ \qquad z^{3}\rightarrow \tfrac{1}{\sqrt{2}} \ , \ \qquad  z^{4}\rightarrow \tfrac{1}{\sqrt{2}} \ .
\end{align}

The second special solution of \rf{310}--\rf{312} (that was already found  in 
\cite{Bandres:2009kw}) 
 is a \lon one for which $\J$ and thus $J$  is not bounded.  Here (cf. \rf{s8},\rf{313}--\rf{s9})
\begin{align}
&r_1=r_2=r_3=r_4=\ha   \ , \qquad    m_1=m_2=-m_3=-m_4 \equiv \ha m  \ ,  \qquad 
w_1=-w_2=w_3=-w_4= \J \ ,  \la{320}\\
& A_0=A_1=0 \ ,  \qquad  \ka^2 = 4\mathcal{J}^2+m^2 \ ,\qquad \   
g_{\a\b}= \bc^2 \eta_{\a\b} \ ,\qquad \bc^2 =  \tfrac{1}{4} m^2 \ ,  \qquad  \Lambda=-\J^2 + \tfrac{1}{4} m^2 \ ,  \la{3200}  \\
& z_{1}=\tfrac{1}{2}\, e^{i(\mathcal{J}\xi^{0}+ \frac{1}{2}m \xi^{1})} \ ,
\ \ \   z_{2}=\tfrac{1}{2}\, e^{i(-\mathcal{J}\xi^{0}+\frac{1}{2}m\xi^{1})}\ , \ \ \  
z_{3}=\tfrac{1}{2}\, e^{i(\mathcal{J}\xi^{0}-\frac{1}{2}m\xi^{1})}\ , \ \ \ 
{z}_{4}=\tfrac{1}{2}\, e^{- i(\mathcal{J}\xi^{0}+\frac{1}{2}m \xi^{1})} \ ,  \la{321}\\
&
\chi=\tfrac{\pi}{4} \ , \ \ \  \ \  \ \ \ \theta_{1}=\theta_{2}=\tfrac{\pi}{2} \ , \ \ \ \ \ \ \ \  \psi = m \xi^{1} \ , \ \ \ \ 
\ \ \ \phi_{1}=\phi_{2}=2\mathcal{J}\xi^{0}\ ,  \la{322}\\
& \la{s10} 
J_3=0 \ , \qquad J_{1}=J_{2} = J=  \sqbl\, \mathcal{J} \ , \qquad 
\ \ \ E_0 =\sqbl \, \kappa =\sqbl \,\sqrt{4{\J}^2+m^2}  = \sqrt{4{J}^2+m^2\bl}\ .
\end{align}
Like for  the similar \adss  solution \ci{Frolov:2003qc} 
here  the energy   expanded at large $\J$ has a  familiar ``fast-string"  form 
\begin{equation} \la{326}
    E_0 = 2J+\frac{m^2\bar{\lambda}}{4J}-\frac{m^4\bar{\lambda}^2}{64J^3} +... \ .
\end{equation}
Note that the two   solutions \rf{314} and \rf{320}  coincide 
in the special  
case of $a=\ha $   and   $\mathcal{J}=\ha m $  when in   both  cases  $E=m \sqrt{2\bl}$
and $z^a$ in \rf{315},\rf{321}  have $\pm \xi^0 \pm \xi^1$  as their phases.\foot{For comparison,  in the  \adss  case
\ci{Frolov:2003qc}
the \sho solution  written in $S^5$ embedding coordinates is 
$t=\ka \xi^0$, \ \  $ X_1 + i X_2 = a  \, e^{i m (\xi^0 + \xi^1)}, \   X_3 + i X_4 = a  \, e^{i m (\xi^0 - \xi^1)},\  
X_3 + i X_4 = \sqrt{1-2 a^2}$ with $\ka^2 = 4 m^2 a^2= 4\J, \ \  J_1=J_2 = \sql\, \J, \  E= \sqrt { 4 m \sql J}$.
For the  \lon solution 
$X_1 + i X_2 = {1\ov \sqrt 2}   \, e^{i  (\J\xi^0 + m \xi^1)}, \ \  X_3 + i X_4 =  {1\ov \sqrt 2}   \, e^{i (\J \xi^0 - m  \xi^1)},\ \ 
X_3 + i X_4 = 0,$ \  \  $E= \sqrt{ 4 J^2 + m^2 \l}$.}

\subsubsection{M2 brane  solutions}
Let us now discuss  how to ``uplift"  the above string solutions to the   M2 brane   solutions in \adsm  so that the 
 brane wrapped on 11d angle $\vp$  and rotating in $\CP^3$. 

 As is well known,  the ``double dimensional reduction"  relates the  M2  brane action  in 11d supergravity background  \rf{013}--\rf{034}
 to the type IIA 
 string action in the  corresponding 10d background \cite{Duff:1987bx, Achucarro:1989dd,Meissner:2022lso}. 
 Namely,  with  a 10+1 split of the target space coordinates and a 2+1 split of the world volume coordinates  one  assumes that 
\begin{equation}\la{324}
    X^{M} = (X^{\mu}, \vp) \ , \ \ \ \xi^{i}=(\xi^{\alpha}, \xi^{2}) \ , \qquad \quad 
    \vp  = \xi^2 \ ,  \ \ \ \partial_{2}X^{\mu} = 0 \ , \ \ \ \partial_{\vp}G_{MN}=0 \ , \ \ \ \ \  \partial_{\vp} C_{MNP} =0\ ,
\end{equation}
and to get the   string action  keeps only the  zero mode in the Fourier expansion  of the M2 brane fields 
in $\xi^2$. 
In the present case of  the \adssZ background \rf{05}--\rf{7},\rf{A12}    where 
 $\y$ is the isometric coordinate of the  $U(1)_{k}$ fiber of $S^7/\mathbb{Z}_k$  the conditions
  $\partial_{\vp}G_{MN}=  \partial_{\vp} C_{MNP} =0$ are indeed satisfied. 

Considering an M2 brane  located at the center of AdS$_4$  and moving in $S^7/\mathbb{Z}_k$ 
the  bosonic  part of its  action  may be written  like  in \rf{s2} 
in terms of coordinates $z^a$ 
 of $ \mathbb{C}^4/\mathbb{Z}_{k}$  with  the additional constraint $z^a \bz_a=1$  imposed by a Lagrange multiplier:
\begin{equation}\label{m2}
    S = -\T_2 \int d^3 \xi \, \sqrt{-\det g_{ij}} \, \big[1 - \ha  {  \Lambda}(\xi)(\bar{z}_{a}z^{a}-1)\big{]} \ ,
    \qquad  \ \ \ g_{ij}=-\tfrac{1}{4}\partial_{i}t\partial_{j}t + \partial_{(i}\bar{z}_{a}\partial_{j)}z^{a} \ .
\end{equation}
Here $z^{a}\equiv  e^{2\pi i\ov k} z^{a}$ or, equivalently, given by \rf{A1}. 
The effective tension $\T_2$ was defined in \rf{011}.
The  corresponding equations of motion  are  
\begin{align}\label{m3}
      \nabla^2  t = 0 \ ,  \qquad 
          \nabla^2 z^{a} = -{\Lambda} z^{a} \ , \qquad  \nabla^2= \tfrac{1}{\sqrt{-g }}\partial_{i}(\sqrt{-g}g^{ij}\partial_{j}) \ , \ \ \ \ \ 
         \bar{z}_{a}z^{a} = 1 \ . \end{align}
It is  straightforward to check that they
are satisfied by $t=\ka \xi^0$   and   
\begin{equation}\label{m4}
    z^{a}(\xi^{i}) = e^{{i\ov k} \xi^{2}}\, z^{a}(\xi^{\alpha}) \ , \qquad \qquad \xi^{2}\in [0, 2\pi) \ , 
\end{equation}  
where  $z^{a}(\xi^{\alpha})$ solve  the equations  \rf{s3},\rf{s70} 
  for a string in $\mathbb{R}\times \CP^3$.
The induced 3d metric $g_{ij}$ can be written as 
\begin{equation}\label{m5}
    g_{ij} = \begin{pmatrix}
g_{\alpha \beta} + A_{\alpha}A_{\beta} & \frac{1}{k}A_{\alpha} \\
\frac{1}{k}A_{\beta} &\frac{1}{k^2}  
\end{pmatrix} \ , \qquad \qquad g_{\alpha \beta}  = \bc^2 \eta_{\a\b} \ , 
\end{equation}
where 
$A_{\alpha}$ and $g_{\alpha\beta}$ are given by \eqref{s30} and \eqref{s70} respectively.

As in the string case, the action \eqref{m2} is invariant under the 
global $SU(4)$ symmetry and   time $t$ translations. 
In particular, for $z^{a}(\xi^{\alpha})$ satisfying the Virasoro constraints  \eqref{s70}, the expressions for the 
conserved charges can be written as  in \rf{s6}
\begin{align}
    J_{r} = \frac{1}{2 k} \T_2 \int_{0}^{2\pi} d \xi^1 \int_{0}^{2\pi} d \xi^2 \Big{[} (\partial_{0}z)^{\dagger}H_{r}z - z^{\dagger}H_{r}\partial_{0}z\Big{]} \ ,\qquad \qquad 
    E_0 = \frac{1}{4 k} \T_2 \int_{0}^{2\pi}d\xi^{1} \int_{0}^{2\pi}d\xi^{2} \ \partial_{0}t \ . \la{329}
\end{align}
These  coincide  with the corresponding string charges \eqref{s6} as 
$\T= {\pi\ov 2k} \T_2$ (see   \rf{0115}). 

Thus  the M2  brane  counterparts of the 
\sho and \lon  string solutions  are   represented  by \rf{m4}  with $z^{a}(\xi^{\alpha})$  
given by \rf{315} and \rf{321} respectively  and the same values of  spins and energies  as in \rf{s9} and \rf{s10}. 
For these solutions  both $g_{\a\b}$   and $A_\a$ are constant (see \rf{3144},\rf{3200})   so that $g_{ij}$ in \rf{m5}  is also constant
\be \la{330}
ds^2_3=  g_{ij}\, d \xi^i d\xi^j=  \bc^2\,   [ - (d \xi^0)^2 + (d\xi^1)^2 ] + \tfrac{1}{ k^2} \big( d \xi^2 + k A_1 d \xi^1\big)^2 \ .\ee
Note that   while for the \lon solution in \rf{3200}  one has  $\bc^2=\four m^2$   and $A_\a=0$   so that the 3d metric  is diagonal, 
for the \sho one  in \rf{3144}  $\bc^2= {1\ov 8} \ka^2$  and $A_0=0$ but  $A_1= 2m (a^2 - \four)$  is  non-zero 
(and, as already  mentioned  above, 
cannot be,  in general,  eliminated by a redefinition of $\xi^2$).  
As a result, in  the \sho   case $g_{ij}$  in \rf{330} 
 represents  a non-trivial torus in the $(\xi^1,\xi^2)$ directions   
 \begin{align}
  \la{331}
&  ds^2_3=  - \bc^2 ( d \xi^0)^2   + \tfrac{1}{ k^2} \big| d \xi^2 + \tau d \xi^1\big|^2 \ ,\qquad \ \ \  \tau=\tau_1 + i \tau_2 \ ,\qquad 
\bc= \sqrt{\tfrac{ m}{2} \J}  
 \ , \\
&\la{3311}
 \tau_1= k\,  A_1= 2km ( a^2 - \four)= -\ha  k m\sqrt{1 -2 \J} \ , \qquad \qquad 
 \tau_2 = k\, \bc = 2k m a \sqrt{ \ha - a^2} = k \sqrt{ \tfrac{ m}{2} \J} \ . \end{align}
 For the \lon  solution one may also write the diagonal metric   in the  form \rf{331}   where 
 \be \la{3319}
\bc= \ha m \ , \ \ \qquad \ \    \tau= i \tau_2 \ , \ \ \ \qquad   \ \   \tau_2= k\, \bc = \ha k m \ . \ee


\subsection{1-loop correction  to the  energy}

Our aim will be  to compute the 1-loop   corrections  to the energies of the \sho and \lon   M2 brane solutions. 
The first step is to find the  corresponding quadratic fluctuation action that follows from \rf{013}--\rf{034}.
This is  can be done, e.g.,    in the static gauge as  in \cite{Drukker:2020swu,Giombi:2023vzu,Beccaria:2023ujc}
(see also a   discussion  in Appendix \ref{ap1}). 
Like in the case of the long folded  M2  brane  solution  in section 2  the induced metric \rf{330} is constant (cf. \rf{300}) 
as are the derivatives of the background  3d  fields 
 so that the fluctuation  Lagrangian has constant coefficients  and  the  spectrum 
 of fluctuation frequencies  is straightforward to find. 

In particular,   the 8 bosonic   fluctuations propagating in the   induced 3-metric \rf{330},\rf{331}
will be described 
by a  coupled   quadratic  2-derivative action with constant coefficients. 
For a single 3d scalar  field $X(\xi)$  with mass $M$ the corresponding Klein-Gordon  operator will   be (cf. \rf{40}) 
\be
  ( - g^{ij} \del_i \del_j  + M^2) X  \ \  \to \ \ 
 \label{sm1a}
   \bc^{-2} \big{[}\partial^2_{{0}}- (\partial_{{1}}-kA_{1}\partial_{2})^2- k^2 \bc^2\partial^{2}_{{2}} + \bc^2 M^2\big{]} X \  .
\ee
Expanding  in Fourier modes in $\xi^i$  as 
\be \la{332}
X(\xi) = \int { d \om \ov 2\pi} \sum_{n =-\infty}^\infty \sum_{ l =-\infty}^\infty   \, \td X_{nl}(\om ) \  e^{i(\om \xi^{0}+n\xi^{1}+l\xi^{2})}
\ , \ee 
the frequencies  $\omega (n,l)$  corresponding to \rf{sm1a}    may be written as  (cf. \rf{331}) 
\begin{equation}\la{3322}
    \omega^2(n,l) = \big|n-\tau\,  l\big |^2 +\bc^2 M^2 = (n- \tau_1 l)^2 + (\tau_2 l)^2 + \bc^2 M^2 \ .
\end{equation}
Assuming  that   one can diagonalize   the $8\times 8$ 
 matrices   for  the  bosonic and fermionic characteristic frequencies  
one  will then get the  1-loop correction to the AdS$_4$ energy   given by 
\begin{align}\la{333}
    E_1=\frac{1}{2\kappa}\sum_{n=-\infty}^{\infty}\sum_{l=-\infty}^{\infty}\, \Omega(n, l)
    \ , \qquad \ \ \  \Omega(n, l) =\sum_B \omega_{_{\text{B}}}(n, l) - \sum_F \omega_{_{\text{F}}}(n, l) \ , 
\end{align}
where $ \Omega(n, l) $ depends on the  parameters of a solution, i.e. $\J$ and $m$. 
The sum over $l$ can be split as in \rf{1555}  into the $l=0$ (string)  contribution 
 and that of the rest of the M2 brane  $l\not=0$  (``KK" ) modes, i.e.   
\begin{align}\la{3333}
    &E_1=E_{1,\text{str}} +E_{1,\text{kk}}  \ , \qquad \qquad 
    E_{1,\text{str}} =\frac{1}{2\kappa}\sum_{n=-\infty}^{\infty}\Omega(n, 0) \ , \qquad   E_{1,\text{kk}} =\frac{1}{2\kappa}\sum_{n=-\infty}^{\infty}\sum_{l\neq 0}\Omega(n, l) \ .
\end{align}
In practice, finding  the explicit   expressions   for the frequencies 
$\omega_{_{\text{B}}}(n, l) $  and $  \omega_{_{\text{F}}}(n, l) $ and thus $\Omega(n, l)$   is hard 
 due to non-trivial mixing of  the transverse fluctuations (cf. Appendix \ref{ap3}).   
 One can use instead  an equivalent 
  representation for $E_1$ in terms of the 1-loop   partition function (cf. \rf{15}  and a discussion 
in \ci{Beccaria:2012xm})\foot{In general,   in  the $l=0$ string case   the contributions of 
 some low-$n$ modes may require special treatment.}
\begin{equation}\label{sm2}
    E_1
     = \frac{1}{2\kappa}\sum_{n=-\infty}^{\infty}\sum_{l=-\infty}^{\infty} \int_{- \infty}^{\infty}\frac{dw}{2\pi}\ \log \frac{\OD_{\text{B}}(w^2, \tau, \mathcal{J})}{\OD_{\text{F}}(w^2, \tau, \mathcal{J})} \ .
\end{equation}
Here  $\OD_{\rm B,F}$ are the determinants of the quadratic fluctuation matrices for
the  bosons and fermions
obtained after expanding in the Fourier modes  as  in \rf{332}
and $w=i\omega$.

\subsubsection{``Short" M2  brane}

Below   we will consider the case of the minimal winding number $m=1$  (corresponding to the 
state   with minimal energy for  given spins). 
Let us first  discuss   the contribution  of the string ($l=0$) modes. 
For the 8=1+3+2+2 bosonic modes  one finds the following expressions for the characteristic frequencies
(see Appendix \ref{ap2})
\begin{align}
 & l=0:\qquad     \omega^2 = n^2 \ , \ \ \qquad \ \ \omega^2=n^2 + 4\mathcal{J} \ \ \ (\text{3\, modes})
   \ , \no \\
    &\omega^2 = n^2 +2 - 3\mathcal{J} \pm \sqrt{\mathcal{J}^2+4 n^2 -4\mathcal{J}n^2} + 2 \sqrt{ (1-2 \mathcal{J}) \big(1+n^2-\mathcal{J}\pm \sqrt{\mathcal{J}^2 +4 n^2 -4\mathcal{J}n^2}\big)} \ , \no \\
    &\omega^2 = n^2+2  - 3\mathcal{J}\pm \sqrt{\mathcal{J}^2+4 n^2 -4\mathcal{J}n^2} - 2 \sqrt{(1-2 \mathcal{J}) \big(1+n^2-\mathcal{J}\pm \sqrt{\mathcal{J}^2 +4 n^2 -4\mathcal{J}n^2}\big)} \ .\label{334} 
\end{align}
The   $8= 2 \times 2+ 2\times 2$  fermionic $l=0$  frequencies are 
\begin{align}
    l=0: \qquad 
    & \omega^2 = 1+n^2+\mathcal{J} \pm 2 \sqrt{ \mathcal{J} +n^2-\mathcal{J} n^2} \ \ \ \ \ \  \ \ \ (\text{2\, modes}) \ , \no \\
&      \omega^2 = 1 +n^2-\mathcal{J} \pm 2 \sqrt{(1-2 \mathcal{J}) \left(\mathcal{J} +n^2\right)} \ \ \ \ \  (\text{2\, modes}) \ ,\label{335}
\end{align}
where 
$\J =8 a^2(\tfrac{1}{2}-a^2)$, $\ka^2= 4 \J$   (see \rf{s9} with $m=1$).
Separating the   special $n=0,1,2$   modes,  
 $E_{1,\text{str}} $ in \rf{3333} may be written as 
\begin{equation}\label{sm1}
  E_{1,\text{str}} = \frac{1}{2\sqrt{\mathcal{J}}}\Big{[} \tfrac{1}{2}\Omega(0, 0) + \Omega(1, 0)+ \Omega(2, 0) + \sum_{n=3}^{\infty}\Omega(n, 0)\Big{]} \ , 
\end{equation}
where $\Omega(n, 0)$ is the  total contribution of the bosonic   and fermionic modes as in \rf{333}. 
Expanding in small $\J$  we get 
\begin{align}
    &\la{337}
    \te \Omega(0, 0) = -4+6 \sqrt{\mathcal{J}}+2 \mathcal{J}+\mathcal{J}^2+\mathcal{O}\big(\mathcal{J}^{5/2}\big) \ , \ \ \ \Omega(1, 0) = 2-2 \sqrt{\mathcal{J}}+\frac{1}{2}\mathcal{J}-\frac{211 }{32}\mathcal{J}^2+\OO\big(\mathcal{J}^{5/2}\big) \ , \\
    &\no \te \Omega(2, 0) = -\frac{1}{3}\mathcal{J}+\frac{131}{216} \mathcal{J}^2+\OO\big(\mathcal{J}^{5/2}\big) \ , \ \ \ \frac{1}{2}\Omega(0, 0) + \Omega(1, 0)+ \Omega(2, 0) =\sqrt{\mathcal{J}}+\frac{7 }{6}\mathcal{J}-\frac{4741 }{864}\mathcal{J}^2+ \mathcal{O}\big(\mathcal{J}^{5/2}\big)  , \\ 
    &\qquad \qquad \qquad\qquad  \sum_{n=3}^{\infty}\Omega(n, 0) = q_{1}\mathcal{J} + q_{2}\mathcal{J}^2+ \mathcal{O}\big(\mathcal{J}^{5/2}\big) \ , \la{338}\\ 
    &\qquad \qquad q_{1} = -\sum_{n=3}^{\infty}\tfrac{2}{n(n^2-1)}=-\tfrac{1}{6} \ , \ \ \ \qquad   q_{2} = \sum_{n=3}^{\infty}\tfrac{ 23 n^4-29 n^2+10}{2 n^3 \left(n^2-1\right)^3} = \tfrac{4741}{864}-\tfrac{9}{2}\zeta(3) \ .\la{339}
\end{align}
As a result,  
\begin{equation}\label{sm7a}
     E_{1,\text{str}}  = \tfrac{1}{2}+\tfrac{1}{2}\sqrt{\mathcal{J}}-\tfrac{9}{4}\zeta(3)\mathcal{J}^{3/2}+ \mathcal{O}\left(\mathcal{J}^{2}\right) = \tfrac{1}{2}+\tfrac{1}{2}\frac{\sqrt{J}}{\bar{\lambda}^{1/4}}-\tfrac{9}{4}\zeta(3) \frac{J^{3/2}}{\bar{\lambda}^{3/4}}+\mathcal{O}\big(\frac{{J}^{2}}{\bar{\lambda}}\big) \ ,
\end{equation}
Combined  with the classical contribution in \eqref{s9} this gives 
\begin{equation}\label{sm79}
    E_{\text{str}}= 2\sqrt{\sqbl J} 
    +\tfrac{1}{2}+\tfrac{1}{2}\frac{J^{1/2}}{\bar{\lambda}^{1/4}}
    -\tfrac{9}{4}\zeta(3) \frac{J^{3/2}}{\bar{\lambda}^{3/4}}+\mathcal{O}\big(\frac{{J}^{2}}{\bar{\lambda}}\big) \ ,
\end{equation}
which   has   similar  structure as  the corresponding  expression in the \adss case  \ci{Roiban:2009aa}.

\

Let us now   consider the $l\not=0$ (membrane-mode)   contribution $E_{1,\text{kk}}$     to the  1-loop energy in \rf{3333}.
We will be interested in its expansion first at large $k$   and then  in   small $\J$. 
 If  the  small $\J$  limit   is   taken  before the large $k$  one   directly 
  in  $\Omega(n,l)$, i.e. before summing over $n,l$, 
this  leads to inconsistencies,  since
  the frequency  lattice   in \rf{3322}
   becomes degenerate  as  $\tau_2\sim \J \to 0 $ (cf.  \rf{3311}).\foot{One can draw  some analogy  
   with  what one finds  for  the non-holomorphic Eisenstein series $E(s, \tau)$ as a function of $\tau$. If one considers its Fourier expansion, it can be seen that the series has a regular behaviour for  $\tau_2 \rightarrow \infty$, while  for 
   $\tau_2 \rightarrow 0$ it is  divergent because of the asymptotics of  the modified Bessel function $K_{\nu}(x)\sim  {x^{-\nu}}$ near zero.}
   Thus it is important that  the large $k$ limit is to  be taken  before   the 
   small $\J$  one.\foot{As discussed in section 1.4  this 
   is consistent with  the standard  't Hooft   large $N$  expansion on the gauge theory  side where one 
      should  first take $N$ 
    large and then  consider limits  of small or large $\l$   and  small  or large  ${J\ov \sql }$.}
     This    implies that 
   \begin{equation} k\gg 1, \ \ \ \  \J \ll 1 \  : \ \ \ \ \ \ \ \ \ \ \ \ \ \   \tau_{2} = \tfrac{1}{\sqrt2}  k \sqrt{\mathcal{J}} \gg 1 \ .\la{3501}  \end{equation}
Below  we  shall  use  the  integral representation \rf{sm2}  for the 1-loop  energy 
and
treat $\tau$ and $\mathcal{J}$ as independent parameters, assuming that  $\tau_{2} \gg 1$ and $\mathcal{J}\ll 1$.
 We will  replace   $\tau$   with its explicit value in \rf{3311}  at the end of the calculation. 
  
 Let us expand the integrand in \rf{sm2} as 
  \begin{equation}\la{351}
    \log \frac{\OD_{\text{B}}(w^2, \tau, \mathcal{J})}{\OD_{\text{F}}(w^2, \tau, \mathcal{J})} \Big|_{\J\to 0} 
    = \EE_{0}(w^2, \tau) + \sqrt{\mathcal{J}} \EE_{1}(w^2, \tau) + \mathcal{J}\EE_{2}(w^2, \tau) + ...
\end{equation}
Using the expressions in  Appendix \ref{ap2}  one finds
that the expressions for the determinants $\OD_{\text{B}}$ and $\OD_{\text{F}}$  depend on $\tau$ 
only through $p^2$ and $q$  defined as 
\begin{equation}
     p^2 \equiv  |n-\tau l |^2 =q^2 +(\tau_{2} l)^2  \ , \qquad \qquad \qquad   q \equiv  n-\tau_{1}l \ . \la{354}
\end{equation}
Explicitly,  we get\foot{To get the expansion in terms of $\mathcal{J}$, we have assumed that $a<\frac{1}{2}$ and  used \rf{s9} to express $a$ in terms of $\J$,  i.e.  $a^2 = \frac{1}{4}-\frac{1}{4}\sqrt{1-2\mathcal{J}}$.} 
\begin{align}
    &\EE_{0} =\te \log \frac{\left(p^2+w^2\right)^5 \big[p^6+3 p^4 w^2+p^2 \left(-8 q^2+3 w^4+8 w^2-16\right)-8 q^2 \left(w^2-4\right)+w^2 \left(w^2+4\right)^2\big]}{\big[p^4+2 p^2 w^2-2 q^2+\left(w^2+1\right)^2\big]^4}   \ ,\la{352} \\
    &\EE_{1} = \te\frac{8 \sqrt{2} q \sqrt{p^2-q^2} \big[4 p^4+p^2 \left(-6 q^2+14 w^2-17\right)-6 q^2 \left(w^2-4\right)+10 w^4+23 w^2+4\big]}{\big[p^4+2 p^2 w^2-2 q^2+\left(w^2+1\right)^2\big] \big[p^6+3 p^4 w^2+p^2 \left(-8 q^2+3 w^4+8 w^2-16\right)-8 q^2 \left(w^2-4\right)+w^2 \left(w^2+4\right)^2\big]} \ .\la{353}
\end{align}
It is also straightforward to find  $\EE_2$ but  its expression is somewhat long  so we will not present it here.
Let us define 
\begin{equation}\la{355}
    \E_{r}(\tau) = \sum_{n=-\infty}^{\infty}\sum_{l=-\infty}^{\infty} \int_{0}^{\infty}{dw\ov 2\pi} \  \EE_{r}(w^2,p^2,q) \ .
\end{equation}
Then combining \rf{sm2},\rf{351},\rf{355} we get 
\begin{equation}\label{sm8}
    E_1= \frac{1}{\kappa}\sum_{r=0}^{\infty}\E_{r}(\tau)\, \mathcal{J}^{r/2}  =  \frac{1}{2}\sum_{r=0}^{\infty}\E_{r}(\tau)\, \mathcal{J}^{(r-1)/2} \ ,
\end{equation}
where we  used  that according to \rf{s9}  $\kappa = 2 \sqrt{\mathcal{J}}$. 
 
 To evaluate $\E_r(\tau)$ in \rf{355}  we   may  assume  that $\EE_{r }$  with an even $r$ 
   is  an even  function of $q$  while  
  $\EE_{r}$  with an odd $r$  
    is an odd function of $q$ (we checked this  property 
  for low   values of $r=0,1,2$ that we will consider below). 
 Then  the  sum  of odd $\EE_r$ over $(n,l)$ in \rf{355}   is zero, since the  terms with 
  $(n, l)$ and $(-n,-l)$ contribute with an opposite sign. 
  Thus we  may   consider only  $\E_r(\tau)$  with an even $r$. 
  One can further use that since the dependence of $\EE_{r}$ on $\tau$   is only  via $p^2$ and $q$, 
  they  are periodic functions of $\tau_1$ and thus  can be expressed in Fourier series    as 
\begin{align}
  &  \E_r(\tau_1+1,\tau_2) = \E_r(\tau_1,\tau_2) \ , \qquad \qquad 
    \E_r = \sum_{s = -\infty}^{\infty}e_r^{(s)}(\tau_{2})\ e^{2\pi i s \tau_{1}} \ , \label{sm4a} \\
    &e_r^{(s)}(\tau_{2}) =\sum_{n=-\infty}^{\infty}\sum_{l=-\infty}^{\infty} \int_0^{\infty}  \frac{dw}{2\pi} \int_{0}^{1}d\tau'_{1}\, e^{-2\pi i  s\tau'_{1}}\, \EE_r(w^2, p'^2, q') \ , \label{sm4}
\end{align}
where $p'$ and $q'$  are   assumed to depend on $\tau_1'$. 

Let us first  consider   the $l=0$ term  in the sum (which  should be the string-theory  contribution already discussed above). 
  In this case from \rf{354} we  have  $q^2 = p^2 =n^2$, i.e. do not depend on $\tau$ 
  and  thus the only non-vanishing term in \eqref{sm4} is the one   with $s=0$, i.e.\footnote{As was  already alluded to 
  above, 
   the representation of the 1-loop correction to energy \rf{333} in terms of the integral in \rf{sm2} 
   is valid when  the integrand in  \rf{sm2} 
    does not have branch points on the real axis as a function of $w$. 
    As this  is not  true  in the special  cases when  $l=0$ and $ \ n =0, \pm 1, \pm 2$, 
    we  are to use instead the representation \rf{337} for these contributions. 
    For the other values of $n$  the  results following from \eqref{359}   and   \eqref{338}   coincide.} 
\begin{equation}
     (\E_{1,{\rm str}})_r=
     e_r^{(0)}(0)  =
     \sum_{n=-\infty}^{\infty} \int_{0}^{\infty}{dw\ov 2 \pi} \  \EE_r(w^2, n^2, n) \ .\la{359}
\end{equation}
The remaining sum over $n$ and $l\not=0$ in \rf{sm4} can be written as:
\begin{align} \label{sm5}
   &2 \sum_{l=1}^\infty \sum_{n=-\infty}^{\infty} \int_0^{\infty}  \frac{dw}{2\pi} \int_{0}^{1}d\tau'_{1}\, e^{-2\pi i  s\tau'_{1}}\, \EE_r 
=  2\sum_{l=1}^{\infty}\sum_{n\text{ mod }l} \int_{0}^{\infty} \frac{dw}{2\pi}  \int_{-\infty}^{\infty}d\tau_{1}\, e^{-2\pi i  s\tau'_{1}}\EE_r\big(w^2, |n-\tau' l|^2, n-\tau'_{1}l\big) \nonumber \\
    &\qquad \qquad = 2\sum_{l=1}^{\infty}\sum_{n\text{ mod }l} e^{-2\pi i s n/l}\int_{0}^{\infty}   \frac{dw}{2\pi}  \int_{-\infty}^{\infty}d\tau'_{1}\,  e^{-2\pi i  s\tau'_{1}} \EE_r\big(w^2, (\tau'^2_1 + \tau_2^2) l^2, \tau'_{1}l\big)  \  .
\end{align}
Here we used the assumption that $\EE_r$ is an even function of $q$
and   also  the properties of the sum and the periodicity of the integral over $\tau'_{1}$, and finally 
shifted  $\tau'_{1}\rightarrow \tau'_{1} + n/l$.

The integral over $\tau_1'$ is   hard to evaluate explicitly even for $\EE_0$ in \rf{352}. 
To proceed, we  shall  focus on   the  large $k$ expansion, i.e. assume as in \rf{3501} that $\tau_{2} \gg 1$.
Rescaling the  integration variables $w = \tau_{2} y$ and $ \tau'_{1} = \tau_{2} x$,  we get for   \eqref{sm5}
\begin{equation}\label{sm5b}
    2\tau_{2}^2\sum_{l=1}^{\infty}\sum_{n\text{ mod }l} e^{-2\pi i s n/l}\int_{0}^{\infty}{dy\ov 2 \pi}  \int_{-\infty}^{\infty}dx \ e^{-2\pi i  s\tau_{2}x} \ \EE_r\big{(}\tau_{2}^2(x^2+1)l^2, \tau_{2}xl, \tau_{2}^2y^2\big{)} \ .
\end{equation}
We first note that if $\EE_r$ is an integrable function of $x$, the integral over $x$ vanishes in the limit of 
$\tau_{2}\rightarrow \infty$ if $s \neq 0$ due to the Riemann-Lebesgue lemma. This  suggests
 that  for $\tau_{2} \gg 1$  the contribution $e^{(s\neq 0)}_r$ in \rf{sm4}  will be suppressed 
relative to  $e^{(0)}_r$  in \rf{359}.\foot{For instance,  in the case of  non-holomorphic Eisenstein series, such terms are exponentially suppressed. This also  happens  for  the  integral  
 $   \int_{-\infty}^{\infty}dx \ e^{-2\pi i s \tau_{2}x}\ \frac{f(x)}{(x^2+x_{0}^2)^{\ell}} \  \sim \ e^{-2\pi s\tau_{2}|x_{0}|}$
  , $   \tau_{2} \gg 1\ $, 
where $f(x)$ is a polynomial with degree less than 
 $\ell \in \mathbb{Z}_{+}$ and  has no poles at $x_{0} \in \mathbb{R}$.}

  If 
the same is true  for all terms in   \eqref{sm5b}, i.e.  
 the terms with $s\neq 0$ are exponentially suppressed,
 then 
  the leading-order terms in the expression for $\E_r(\tau)$ in \rf{sm4a}
   can be written as  $(\E_{1,{ \rm str}})_r$  in \rf{359}  plus  $ \big{(} \E_{1,\text{ kk}}\big{)}_r$, 
   i.e.
   \be    \E_r(\tau) = (\E_{1,{\rm str}})_r +  \big{(} \E_{1,\text{kk}}\big{)}_r \ ,    \qquad \ \ \  
    \big{(} \E_{1,\text{kk}}\big{)}_r 
    =  2{\tau_{2}^2} 
    \sum_{l=1}^{\infty} l \int_{0}^{\infty}{dy\ov 2 \pi}  \int_{-\infty}^{\infty}dx \Big{(}  \frac{\hat \EE^{(4)}_r}{\tau_{2}^4}  + \frac{\hat \EE^{(6)}_r}{\tau_{2}^6}+... \Big{)} + ... \ .\la{364}
\ee
Here we have  assumed that $\EE_{r}$ admits  the  large $\tau_2$ 
expansion  of the form $\sum^\infty_{m=4} {\hat \EE^{(m)}_r}\, \tau_{2}^{-m} $, 
where $ {\hat \EE^{(m)}_r} = {\hat \EE^{(m)}_r}   \big{(}\tau_{2}^2(x^2+1)l^2, \tau_{2}xl, \tau_{2}^2y^2\big{)}    $ as in \rf{sm5b}.
 \ We have  checked explicitly  that this is true  for $r=0$ and 2. 


 We then arrive at the following expressions for $\big{(} \E_{1,\text{kk}}\big{)}_{0}$ and  $\big{(} \E_{1,\text{kk}}\big{)}_{2}$:
\begin{align}\label{sm6}
     &\big{(} \E_{1,\text{kk}}\big{)}_{0} = -\frac{4 \zeta(2)}{\tau_{2}^2} - \frac{152\zeta(6)}{15 \tau_{2}^6} + ... \ , \qquad \qquad 
     \big{(} \E_{1,\text{kk}}\big{)}_{2} =  \frac{8 \zeta(2)}{\tau_{2}^2} + \frac{10 \zeta(4)}{\tau_{2}^4}+ ... \ .
\end{align}
Using \rf{3501}, i.e.  that $\tau_{2}^2 = \ha k^2 \mathcal{J}$,  and  plugging \rf{sm6} into  the expansion in \rf{sm8} 
we conclude that the membrane-mode  contribution 
to the 1-loop energy of the \sho solution   can  be written as
\begin{align}\label{sm11}
     E_{1,\text{kk}}
     &= \mathcal{J}^{-1/2}\big{(}-\frac{4\zeta(2)}{k^2\mathcal{J}}+... \big{)} 
     + \mathcal{J}^{1/2}\big{(}\frac{8\zeta(2)}{k^2\mathcal{J}}+... \big{)} +...+ 
      \OO( {1\ov k^4})  
    \ . \end{align}
Combining  this with the  classical  part of the   energy in \rf{s9}  and the string ($l=0$) 
  1-loop  contribution in \rf{sm7a} or \rf{sm7}  we then get   the following  prediction for the 1-loop corrected 
  \sho  M2  brane energy 
\begin{align}
    E_{_{\rm M2} }= &2\sqrt{\sqbl J} 
    +\tfrac{1}{2}+\tfrac{1}{2} \bar{\lambda}^{-1/4} J^{1/2}
    -\tfrac{9}{4}\zeta(3) \bar{\lambda}^{- 3/4}  J^{3/2}+\mathcal{O}\big(\bl^{-1} {J}^{2}\big) \no\\
 & +   {1\ov k^2} \Big[ \zeta(2)  \big(-4 \bl^{3/4} J^{-3/2}   + 8   \bl^{1/4} J^{- 1/2}  \big) + \OO\big(\bl^{-1/4} J^{1/2} \big)\Big]  + \OO( {1\ov k^4}) 
    \ . \label{sm777} \end{align}
    Note that 
like in the  fast-spinning M2 brane  case considered  in section 2 (cf. \rf{22},\rf{211}), 
here the leading $1\ov k^2$ correction is also  proportional to $\zeta(2)={\pi^2\ov 6}$.   
  On   the dual ABJM gauge theory side  \rf{sm777} 
  with  $\bl = 2\pi^2 \l$ and ${1\ov k^2} = { \l^2\ov N^2}$ (see \rf{010}) 
  is a  prediction 
   for the    leading non-planar correction 
  to  the  dimension  of the corresponding   \sho 
  operator.\foot{As 
  was mentioned  above  (cf.  \rf{319}), the \sho    string or \sho  M2  brane solution  has a direct analog   in $\mathbb R^{1,9} \times  S^1$  flat space. 
  There   the  circular M2  brane  is rotating with $J_1=J_2$ in two orthogonal planes in $\mathbb R^4 \subset \mathbb R^{1,9}$ and is wrapped on $S^1$ of radius $\rod$. 
 To   take the flat space limit we need to 
   identify the radius of $S^1$  as $\rod  = {\LL\ov k}$  that
    will be fixed in the large $\LL\sim L $ 
     limit  along with the parameters $\ka$ and $a$  
    of the  solution in \rf{319} (cf. \rf{314}). 
    To get the    energy   and spin   in  the flat space  limit and relate to string theory  we need  to  rescale 
    $E\to   \ha L  E$   so that  $E$ and $J$   will have canonical mass dimensions (1  and 0).
       Then  using   that  $\sqbl = { L^2\ov 4 \a'}$
    (cf. \rf{09},\rf{010})
    we  conclude  that all string corrections to the classical term $E = 2 \sqrt{ \a'^{-1} J}$
      in the  first line of \rf{sm777}
    vanish, in  agreement with the fact that the  free superstring spectrum 
    in flat space is not deformed by $\alpha'$ corrections. 
    At the same time, the  $\rod^2\sim \gs^2$ dependent corrections in the  second line of \rf{sm777}
    survive,   with the leading  one proportional to $ \zeta(2) \alpha'^{-1/2} \gs^2 J^{-3/2}$, i.e. 
    we get 
    $E=  2 \sqrt{ \a'^{-1} J}\big[1 -  {1\ov 2} \zeta(2) \gs^2 J^{-2}  + ... \big]$. 
    This can be  checked   
    1-loop  computation   of the energy of   the \sho M2  solution directly  in the  flat space
    case   and may be related to the expectation that masses of  massive 
    superstring  states   may received 1-loop (and higher order)  corrections 
    (cf. \ci{Sundborg:1988ai,Amano:1988ht,Chialva:2003hg,Chialva:2004xm,Sen:2016gqt}).
  Note that a   non-zero  1-loop correction to the 
  energy   of a different $J_1=J_2$   supermembrane  solution  in   flat space  
  (where the membrane   was rotating in 2 planes with the ``radii''    being periodic functions of $\s^1$ and $\s^2$
  but   was  not wrapped on $S^1$)  was  found also  in \ci{Mezincescu:1987kj}. 
    }


   %
\subsubsection{``Long" M2  brane}

Let us now  consider a similar computation of 1-loop  correction to the energy of the $m=1$ \lon   M2 brane solution 
that generalizes the string solution \rf{320}--\rf{s10}. Here one has    diagonal induced 3-metric as in \rf{300}
(cf. \rf{331}) 
\be \la{367} 
A_\a=0\ , \ \ \ \ \ \ka^2 = 4 \J^2  +1 \ , \qquad \qquad
  ds^2_3 = \four\big[  - (d\xi^0)^2 + (d\xi^1)^2  + \tfrac{4}{ k^2} (d\xi^2)^2\big] \ . \ee
The characteristic frequency polynomials  for the  \lon   solution are given in Appendix \ref{ap3}. 

Like in  the \sho  case let us  first consider the  string theory  $l=0$ contribution. 
We find  as in  \cite{Bandres:2009kw}   that there are $8=1+3 + 2 \times  2$  bosonic  
\begin{align}\label{lm1}
 l=0: \ \ \  \ \ \   &\omega^2 = n^2 + 4\mathcal{J}^2-1 \ , \qquad  \ \ \omega^2 = n^2 + 4\mathcal{J}^2 + 1 \ \  \  (\text{3 modes})  \ , \nonumber \\
    &\omega^2 = n^2 + 2\mathcal{J}^2 \pm \sqrt{4\mathcal{J}^4+(4\mathcal{J}^2+1)n^2} \ \ \  \ \ \ \  (\text{$2\times 2$ modes})  \ , 
\end{align}
and $8=2 \times 2 + 4 $ fermionic  fluctuation frequencies 
\begin{align}\label{lm2}
  l=0: \ \  \qquad  \te  &  \omega^2 = n^2 + 5\mathcal{J}^2 +\tfrac{1}{4} \pm \sqrt{(4\mathcal{J}^2+1)(4\mathcal{J}^2+n^2)}\  \ \ \ \ \   (\text{$2\times 2$  modes}) , \no\\
 & \omega^2 = n^2 + \mathcal{J}^2 + \tfrac{1}{4}  \ \ \ \ \   (\text{4 modes}) \ . 
\end{align}
 These frequencies agree with \rf{334},\rf{335}  for   $\mathcal{J}=\frac{1}{2}$
when the \sho and \lon solutions  become equivalent. 
Note that in contrast to what happens in the \adss case (where the  $m=1$  solution   is unstable \ci{Frolov:2003tu}) 
these   frequencies are always real, i.e. the $J_1=J_2$ solution in \adsc\  is stable for   any $\J$.  
 
The 1-loop energy  string energy  $    E_{1,\text{str}} $ 
is   given  again by the  general expressions in \rf{333},\rf{3333}.
Here we will be interested in its expansion in  the large spin limit   $\mathcal{J} \gg 1$.
 Like in the   \adss case \cite{Beisert:2005cw,Schafer-Nameki:2005aui,Minahan:2005qj},  in addition to  the ``analytic"  contributions 
  (with even powers of $\mathcal{J}^{-1}$) 
 discussed  already in \cite{Bandres:2009kw},  there are also  ``non-analytic" terms
 (with odd  powers of $\mathcal{J}^{-1}$), i.e.   for large $\J$
\begin{equation}\la{370}
   E_{1,\text{str}}  =  \frac{1}{2\kappa}\sum_{n=-\infty}^{\infty}\Omega(n, 0;\mathcal{J}) = E_1^{\text{an}}+E_1^{\text{non}}+\mathcal{O}(e^{-\mathcal{J}}) \ .
\end{equation}
To sum up the series over $n$  we apply  the Abel-Plana summation formula (with a slight modification due to an  additional branch cut coming from the bosonic modes).\foot{To get $ E_1^{\text{non}}$ in    \cite{Schafer-Nameki:2006dtt} an alternative 
 method using the Sommerfeld-Watson transform  was applied.}
 As a result (here $\ka=\sqrt{4 \J^2 +1}$)
\begin{align}
    E_1^{\text{an}}=&\frac{i}{\kappa}\int_{0}^{1}ds\cot(\pi s)\Big{[} \sqrt{4\mathcal{J}^2+1+(s-i\sqrt{1-s^2})^2}-\sqrt{4\mathcal{J}^2+1+(s+i\sqrt{1-s^2})^2}\Big{]}   \label{201} \\
   = & \frac{1}{2\mathcal{J}^2}\Big{[}\tfrac{1}{4}+\sum_{n=1}^{\infty}\big(n\sqrt{n^2-1}-n^2+\tfrac{1}{2}\big)\Big{]} -\frac{1}{8\mathcal{J}^4}\Big{[}\tfrac{3}{16}+\sum_{n=1}^{\infty}\big(\tfrac{3}{8}-n^4+n\sqrt{n^2-1}(\tfrac{1}{2}+n^2)\big)\Big{]} + \mathcal{O}\big{(}\frac{1}{\mathcal{J}^6}\big{)} \ , \no \\
\label{lm3}
     E_1^{\text{non}} =&  \frac{1}{\kappa}\Big{[}2\int_{0}^{\infty}ds\sqrt{2\mathcal{J}^2+s^2+\sqrt{4\mathcal{J}^4+s^2(4\mathcal{J}^2+1)}}+ 2\int_{1}^{\infty}ds\sqrt{2\mathcal{J}^2+s^2-\sqrt{4\mathcal{J}^4+s^2(4\mathcal{J}^2+1)}}  \nonumber\\
 & \ \ \ + \int_{0}^{\infty}dt\Big{(}3\sqrt{s^2+4\mathcal{J}^2+1}+\sqrt{s^2+4\mathcal{J}^2-1}-4\sqrt{s^2+\mathcal{J}^2+\tfrac{1}{4}}-4\sqrt{s^2+4\mathcal{J}^2}\, \Big{)}\Big{]} \ .
\end{align}
The sums that appear  in \rf{201} converge, 
and, in particular, the coefficient of the leading $1\ov \J^2$ term is the same as in 
\cite{Bandres:2009kw}\foot{The  $1\ov \J^2$   correction is  essentially the same as  
in the \adss  case  \ci{Beisert:2005mq} (in general, with the winding numbers related by $m \to \ha m $). } 
\begin{equation}\la{373}
    c_1\equiv  \tfrac{1}{4}+\sum_{n=1}^{\infty} \big(n\sqrt{n^2 - 1} - n^2 + \ha \big) \approx - 0.336\ .  
\end{equation}
Evaluating the integrals in \eqref{lm3} and expanding for $\mathcal{J}$, we get  $1/\J$ and $1/\J^3$ contributions  with $\log 2$ coefficients. Then 
the  combined result for \rf{370}  is 
\begin{equation}\label{lm4}
    E_{1,\text{str}}\Big|_{\J \to \infty}  =  -\frac{\log 2}{2\mathcal{J}} +\frac{c_1}{2\mathcal{J}^2}
     +\frac{\log 2}{16\mathcal{J}^3} + \mathcal{O}(\frac{1}{\mathcal{J}^4}) \ ,
\end{equation}
This is to be added to the large $\J$  expansion of the classical energy in \rf{s10} 
\begin{equation}\la{375}
    E_0 = \bar{\lambda}^{1/2}\sqrt{4\mathcal{J}^2+1} = \bar{\lambda}^{1/2}\Big{[}2\mathcal{J}+\frac{1}{4 \mathcal{J}}-\frac{1}{64 \mathcal{J}^3} + \OO(\frac{1}{\mathcal{J}^5})  \Big{]}\ . 
\end{equation}
As a result, the 1-loop  string energy can be put into the form 
\begin{equation}\label{lm7}
    E_{\text{str}}\Big|_{\J \to \infty} = 2J + \frac{\bar{h}^2(\bar{\lambda})}{4J} + c_1 \frac{ \bar{\lambda}}{2J^2}  -  
    \frac{\bar{h}^4(\bar{\lambda})}{64J^3}    + ... \ ,
\end{equation}
where  $J = \bar{\lambda}^{1/2}\mathcal{J}$    and  $\bar{h}(\bar{\lambda})\equiv  2\pi h(\l) = \bar{\lambda}^{1/2}-\log 2 +...$
with $h(\l)$ that appeared   in \rf{88},\rf{91}. 
 A similar result  that the replacement $\sqbl \to \bar{h}(\bar{\lambda})$ 
happens  only for the  coefficients of the 
 ``odd"  $1/J^{2r+1}$ terms   in the expansion  of the energy (that are then directly related to those in the  \adss  case) 
 was found  in  \cite{McLoughlin:2008he} for a circular rotating string with  spins $S$   and $J$   stretched in both \ads \ and $\CP^3$.


\

Let us now  turn to  the $l\not=0$ membrane mode contribution $E_{1, \text{kk}}$  in \rf{3333}. 
Using the integral representation \rf{sm2}    we   get 
\begin{equation}\label{lm5}
    E_{1, \text{kk}}= \frac{1}{\kappa}\sum_{n=-\infty}^{\infty}\sum_{l \neq 0}\int_{0}^{\infty}\frac{dw}{2\pi}\, \mathcal{E}\big{(}w^2, n^2, (\tau_{2}l)^2, \mathcal{J}\big{)} 
    \ , \ \ \ \ \ \  \E= 
     \log \frac{\OD_{\text{B}}(w^2, \tau_{2}, \mathcal{J})}{\OD_{\text{F}}(w^2, \tau_{2}, \mathcal{J})}\ , \ \ \  \ \  \tau_2 \equiv  \ha k\ . 
\end{equation}
While   the induced  metric in \rf{367}  is diagonal, to keep the analogy   with the \sho  M2 case (cf. \rf{3501},\rf{351}) 
we introduced  as in \rf{3319}  the  coefficient $\tau_2 =  \ha k$    that will  be again large in the $k \gg 1$ limit  we are interested in. 

Here we should  first expand in large $\tau_2$ and then in large $\J$. 
As follows from the explicit form of the  determinants $\OD_{\text{B}}$ and $ \OD_{\text{F}}$ in Appendix \ref{ap3}
the integrand $\mathcal{E}$ in \rf{lm5}  turns out to be an even function of both $n$ and $l$.
  To compute \rf{lm5} in the large $\tau_2$   limit 
     we may  try to follow the same strategy as in the \sho  M2 brane case  discussed  above. 
     For that  we may 
     formally introduce a   parameter $\tau_1$ (to be taken to zero at the end) 
      shifting $n \to n - \tau_1 l$    as  in \rf{351},\rf{354}.  Then we  
     can  take $ l\neq 0$  and consider first  the sum over $n$ following the same steps as in \rf{sm4}--\rf{364}.
     Like   in  the string case, one  may  split the sum into the integral part and  finite series 
      and  the  expectation  is that the  contribution  of the latter is exponentially  will be  suppressed when $k\gg 1$ and $l \neq 0$. 
 This suggests  that like in \rf{364} the sum over $n$ can be effectively replaced by an integral 
 \begin{equation}\label{lm55}
 \tau_2 \gg1: \ \ \ \ \ \  \ \ \ \ \ \ \ 
   E_{1, \text{kk}}\approx  \frac{1}{\kappa}\sum^\infty_{l =1}\int_{0}^{\infty}\frac{dw}{\pi}\,\int^\infty_{-\infty} d n \,  \mathcal{E}\big{(}w^2, n^2, (\tau_{2}l)^2, \mathcal{J}\big{)} \ . 
 \end{equation}
 Assuming that  $\mathcal{J}/\tau_{2} \ll 1$ we may 
 rescale the integration variables in \rf{lm55}  as $w = \tau_{2} y$, $n = \tau_{2}x$
  (cf. \rf{sm5b}) 
 \begin{equation}\la{379}
     E_{1, \text{kk}} =  \frac{\tau_{2}^2}{\pi \kappa}\sum_{l=1}^{\infty}\int_{0}^{\infty}dy\int_{-\infty}^{\infty}dx \  \Big{[} \frac{\mathcal{E}^{(2)}(\mathcal{J})}{\tau_{2}^2} + \frac{\mathcal{E}^{(4)}(\mathcal{J})}{\tau_{2}^4} + ... \Big{]} \ , 
\end{equation}
where we can further expand the integrand   at large $\J$. 
As a result (cf. \rf{sm11})
\begin{align}
    E_{1,\text{kk}} =& \frac{\zeta(2)}{\tau_{2}^2}\big{(}-2\mathcal{J} - \frac{1}{2\mathcal{J}}+\frac{3}{64\mathcal{J}^3} + ... \big{)} + \frac{\zeta(4)}{\tau_{2}^4}\big{(}\frac{21}{4}\mathcal{J}^3+ ... \big{)}+ \OO( {1\ov \tau_2^6}) \nonumber \\
    =&  \frac{4\zeta(2)}{k^2}\big{(}-2\mathcal{J} - \frac{1}{2\mathcal{J}}+\frac{3}{64\mathcal{J}^3} + ... \big{)} + \frac{16\zeta(4)}{k^4}\big{(}\frac{21}{4}\mathcal{J}^3+ ... \big{)}+ \OO({1\ov k^6})\ . \la{380}
\end{align}
Combining this with the string part \eqref{lm7} 
we get (cf. \rf{sm777}) 
\begin{align}
    E_{_{\rm M2}}= 2J  &+ {\bl \ov 4 J} (1 - 2 \log 2\,  \bl^{-1/2} + ...)  +  c_1  {\bl \ov 2 J^2} ( 1 + ...)  + ...\no \\
    &+  {1\ov k^2}\,  \zeta(2)  \big{(}-8\bl^{-1/2} {J} - 2  \frac{\bl^{1/2} }{{J}}+\frac{3\bl^{3/2}}{16{J}^3} + ... \big{)}
    + \OO({1\ov k^4}) \ . \la{3811} \end{align} 
    Here the ${1\ov k^2} = {\bl^2 \ov (2 \pi^2)^2 N^2}$  term represents   the prediction for the  strong-coupling limit of the  leading 
    non-planar correction   to the dimension of  the corresponding operator  with the large  spin $J$.

  \renewcommand{\theequation}{4.\arabic{equation}}
 \setcounter{equation}{0}

\section{Concluding remarks}

In this  paper we discussed  the \adsm M2 brane   counterparts  of the   computations of 1-loop corrections to energies  of the 
three  string solutions in \adsc:  ``long"  folded     string with large   spin
 in AdS$_4$   and \sho and \lon   circular 
 strings  with  equal  angular  momenta $J_1=J_2$   in $\CP^3$. As a result, we  obtained 
  predictions   for the leading non-planar corrections 
 to  scaling  dimensions of the corresponding dual ABJM operators  at strong coupling. 
 
 In all  cases the $1/N^2$ term is proportional to $\zeta(2)={\pi^2\ov 6}$. This  
  is related to the fact that   ${1\ov k}= {\l\ov N}$ is the radius of the 11d circle  $\vp$ which   is identified with the  cylindrical 
 M2  brane  dimension $\xi^2$ so that the dependence  on the  corresponding  Fourier mode  number $l$  is via $k l$. As a result,   the 
  coefficient of the $1\ov k^2$ term is proportional to $\sum_{l=1}^\infty { l^{-2} }= \zeta(2)$.

 There are several  obvious   generalizations. 
 One  may consider the M2  brane   analog of  the folded spinning   string in AdS$_4$ 
  with an extra   orbital  momentum $J$  in $\CP^3$. Taking   the limit when $ S \gg J \gg \sql \gg 1$  with 
  ${\sql \ov  J} \ln {S\ov J}$=fixed    determines  the generalized   cusp anomaly or  
  scaling function.  In  the \adss  string case  this  solution    was  studied  in \ci{Frolov:2006qe,Giombi:2010fa}.
  In this limit the  resulting string   fluctuation  Lagrangian   has constant   coefficients   
    and thus  finding the quantum corrections to the classical 
  energy is  straightforward.\foot{Moreover,  in the  \adss   case  the $(S,J)$  solution in this limit is related by an analytic continuation to the circular  2-spin solution  in $S^5$,  implying  a  relation between the  fluctuation frequencies  \ci{Frolov:2006qe}.}
  For the  string in \adsc\  the 1-loop  correction to the energy  of such $(S,J)$  solution 
  was already  found in \ci{McLoughlin:2008ms,Alday:2008ut}
 and a generalization to the  M2  brane case  in \adsm  should  not be a problem.  
  

One can also    consider  a M2 brane analog  of another $(S,J)$   string 
solution where the    string is wrapped on a circle in both AdS$_4$ and $\CP^3$ part  
(here $S=m J$  where $m$ is a wrapping number).  This is the direct analog of the 
 solution in \adss  studied  in  \ci{Arutyunov:2003za,Park:2005ji,Beisert:2005mq}. 
The 1-loop correction to the energy of this  circular ($S,J)$ string in \adsc\  was computed in  \ci{McLoughlin:2008he}
and a   generalization to the M2  brane case   should be  again straightforward. 
Expanding   in small $S/J$ one  may  relate  \ci{Beccaria:2012xm,Beccaria:2012kp}
 the leading term in the string energy (or in the  dimension of the dual operator) 
 to the so called 
slope function  which,  in the planar limit,   is known exactly   from the 
 integrability    \ci{Basso:2011rs,Gromov:2012eg,Gromov:2014eha}. 

The slope function   turns out to be  very similar to the
 \brf  that can be found  from  localization (via  circular BPS WL connection)   and, in the planar limit, 
 from the  integrability (see   \cite{Correa:2012at, Correa:2012hh,Gromov:2012eu}).
Assuming the analogy between the slope function   and the \brf continues also in the ABJM case, ref. \ci{Gromov:2014eha}
suggested a   conjecture for  the $h(\l)$   function  that enters the  ABJM magnon dispersion relation \rf{91}, and it  
  passed all tests so far.\foot{In  \ci{Gromov:2014eha}  the  comparison was made 
between the structure of the integral representation for the $1\ov 6$ BPS WL   and the ABJM slope function found there.}
It would be interesting to use the above  M2  brane  approach  to find 
a prediction for  non-planar corrections to  the slope  function at strong coupling,  
and then to compare it  to the   known expression  for 
the  \brf $B(\l, N)$ in the ABJM theory 
(see \ci{Lewkowycz:2013laa,Correa:2014aga,Aguilera-Damia:2014bqa,Bianchi:2017svd,Bianchi:2018scb,Bianchi:2018bke,Guerrini:2023rdw,Armanini:2024kww}).
One may also 
 consider  a direct  M2  brane computation of  non-planar corrections to the \brf   following the approach of
 \ci{Giombi:2023vzu} and generalizing to the case of non-trivial wrapping number $\rw$. One may then get the \brf  
  by  taking a derivative over $\rw$ of the large $N$  expansion  of  the log of the  Wilson loop expectation value
  (see a discussion  in Appendix \ref{ap4}).
  
  At a more   conceptual level,  it   would   be remarkable  to find a way to do similar 
  computations  of non-planar   corrections in the type IIB 
  \adss  superstring dual to $\N=4$ SYM theory. 
  While  we  utilized the fact that the type IIA string theory   has an uplift  to   M-theory, allowing to apply the semiclassical M2  brane   approach, there is no   obvious analog of this  procedure   in the type IIB    string theory. 
  At the same time,  the  exact localization  results for the  expectation values 
  of the $1\ov 2$  BPS   Wilson loops in SYM and ABJM   theories  exhibit  very similar  structure when  expanded in $1/N$  \ci{Giombi:2020mhz,Giombi:2023vzu}. 
  Expressing $ \langle W\rangle$   in terms of the string coupling $\gs$ and the string tension $\T$ in the ABJM theory we have (see \rf{0127})
  \be 
  \langle W\rangle_{_{\rm ABJM}}
 =   {\sqrt \T \ov\sqrt{2 \pi} \gs }\, e^{2\pi\,T} \Big[1 
  +{\pi \ov 12}   { g^2_\str \ov \T } +{7  \pi^2\ov 1440}   \big({ g^2_\str \ov \T }\big)^2 +...\Big]
 \ , \ \ \   \T = \sqrt{ \l \ov 2}, \ \ \ 
 \gs= {\sqrt \pi\ov N} ( 2 \l)^{5/4}, \ \ \  \l= {N\ov k} \ .  \la{470} \ee
  In the case of $\N=4$ SYM theory, expressing   
  $ \langle W\rangle$   in \rf{b11}  in terms of  the corresponding $\gs$ 
  and $\T$   gives 
    \be 
  \langle W\rangle_{_{\rm SYM}}
 =   {\sqrt \T \ov 2 \pi \gs }\, e^{2\pi\,T} \Big[1 
  +{\pi \ov 12}   { g^2_\str \ov \T } +{  \pi^2\ov 288}   \big({ g^2_\str \ov \T }\big)^2 +...\Big]
 \ , \qquad
  \T= {\sql\ov 2 \pi} , \ \ \ \  \gs= {g^2_{\rm YM} \ov 4 \pi} , \quad 
 \l= g^2_{\rm YM} N \ . 
   \la{471} \ee
Remarkably, the two  expansions in \rf{470} and \rf{471} 
  have  the same  universal form, and, moreover, the leading  1-loop  $\gs^2$ 
  string correction terms  happen to have  the same  coefficients  \ci{Giombi:2020mhz}.
  
  Surprisingly, the same coefficient of the $ { g^2_\str \ov \T }$ term 
   is  found  also  for  the leading non-planar  correction to the ABJM cusp  anomalous dimension $f(\l, N)$  in \rf{1200},\rf{0129} coming  from the 1-loop  M2  brane contribution  we computed in section 2. Including also the leading string   contributions,    we get from  \rf{045}  and \rf{777}
  \begin{align}
   \la{473}
  f_{_{\rm ABJM}}(\T,\gs)=& {1\ov \pi} \Big[ 2\pi   \T  -  \tfrac{5}{2}  \log 2 +   \OO(\T^{-1})\  + 
  {\pi \ov 12}   { g^2_\str \ov \T } +{  \pi^2\ov 1440}   \big({ g^2_\str \ov \T }\big)^2 +...\Big]
\no\\
 & ={1\ov \pi} \Big[\pi \sqrt{2\l}   - \tfrac{5}{ 2} \log 2 + \OO(\tfrac{1}{ \sql}) +
    {2\pi^2 \ov 3}   { \l^{2}  \ov N^2 } + 
{2\pi^3\ov 45}   { \l^{4}  \ov N^4 }  +...\Big]  \ .  \end{align}
If we make  a  bold conjecture that  the   coincidence of  the  order 
$\gs^2$  string  1-loop coefficients observed in the Wilson  loop expressions in \rf{470} and \rf{471}   should 
extend also to the cusp anomaly,   we may then make a prediction that in the SYM 
theory  the analog  of \rf{473}   should read
\begin{align} \la{4733}
   f_{_{\rm SYM}}(\T,\gs)= &  {1\ov \pi} \Big[ 2\pi   \T  
   - 3\log 2 +    \OO(\T^{-1})  +    {\pi \ov 12}   { g^2_\str \ov \T } +  \g_1 {  \pi^2}   \big({ g^2_\str \ov \T }\big)^2 +...\Big]\no\\
=   & {1\ov \pi} \Big[\sql  - {3} \log 2 + \OO(\tfrac{1}{ \sql}) +
    {1 \ov 12}   { \l^{3/2}  \ov N^2 } + 
{\g_1 \ov 36}   { \l^{3}  \ov N^4 }  +...\Big]  \ . 
 \end{align}
Here  the $1/N^2$  term should be   representing the 
     strong coupling  limit of the leading non-planar correction 
and we introduce $\g_1$ as a    coefficient  of  the subleading non-planar term.
It would be very  interesting to confirm this  prediction that  
 the leading  non-planar correction to the SYM cusp anomalous  dimension 
 that scales  as $\l^4$  at weak coupling (see \rf{508}) 
 should scale as $\l^{3/2}$ at strong coupling.




\section*{Acknowledgments}
We  would like to  thank M. Beccaria,  S. Ekhammar,  L. Guerrini, M. Lagares, S. Penati    and   V. Velizhanin   for useful discussions  and comments.
 The work of SG is supported in part by the US NSF under Grant No.~PHY-2209997.
SAK  acknowledges support of   the President's  PhD Scholarship of Imperial College London. 
AAT is  supported by the STFC grant ST/T000791/1.  
Part of this work was done while AAT was attending 
the   meeting  ``Integrability in low-supersymmetry theories" (Trani,  2024)  funded by the COST Action CA22113, by INFN and by Salento University. 

\np

\small 
\appendix

\renewcommand{\theequation}{A.\arabic{equation}}
 \setcounter{equation}{0}
\section{\la{ap1}  Quadratic fluctuation action}

To find  the 1-loop  correction to the energy one needs to expand the M2 brane action \rf{013}--\rf{034} 
in the \adsm   background  near a  given  classical  solution. 
The spectrum of  bosonic fluctuations  
 will in general contain 8  physical (``transverse")  modes and 3 unphysical  (``longitudinal") 
  modes. The latter   can be   eliminated  by imposing  a static gauge. 
  Alternatively, one can  just isolate 
  the  fluctuations in the normal directions to the surface
    (see, e.g,   \cite{Harvey:1999as,Forini:2015mca,deLeonArdon:2020crs,Goon:2020myi} for  similar discussions). 
    
    Viewing the membrane  as a surface in 11d spacetime, one can define  an orthonormal basis  $e_{i}$ on the membrane
  world volume (here $i,j=0,1,2$  and $A,B$ are tangent-space  11d indices) 
\begin{equation}
    \langle e_{i} , e_{j} \rangle = e_{i}^{A}e_{j}^{B}\eta_{AB} = \eta_{ij} \ .
\end{equation}
For a pair of tangent vector fields $X, Y$  and the Levi-Civita connection $\nabla$  in the target space    one can define 
 the connection $\nabla^{\text{T}}$ on the  brane (corresponding to the induced metric $g_{ij}$)
  and the extrinsic curvature $\rK$  as 
$    \nabla_{X}Y=
     \nabla^{\text{T}}_{X}Y+\rK(X, Y) $. 
 For a   vector $\NN$ in the normal bundle  we also define:
    $ \nabla_{X}\NN  = 
     -\rA_{\NN}(X)+\nabla_{X}^{\perp}\NN $
where $\nabla^{\perp}$ is the connection on the normal bundle and $\rA_{\NN}(X)$ is the Weingarten operator, related to the extrinsic curvature as  $
    \langle \rA_{\NN}(X), Y \rangle = \langle \rK(X, Y), \NN \rangle.$
The bosonic equations of motion for the M2 brane following from \rf{013},\rf{032} 
can  be written as (here $\rK_{ij}=\rK(e_{i}, e_{j})$)
\begin{equation}\label{q1}
    \eta^{ij}(\rK_{ij})_{A}+\frac{1}{3!}\epsilon^{ijk} \, F_{ABCD}E^{B}(e_{i})E^{C}(e_{j})E^{D}(e_{k}) = 0 \ ,
\end{equation}
where $E^A$  is a basis of the  target space 1-forms. 
The quadratic fluctuation part of the  bosonic  M2 brane action   for the fluctuations $\NN$   in the normal directions 
is then ($d^3V = d^3\xi \sqrt{-g}$)
\begin{align} \label{5a}
    S_{B,2} = - T_{2} \int d^3 V \Big{[} \eta^{ij}\langle \nabla^{\perp}_{e_{i}}\NN, \nabla^{\perp}_{e_{j}}\NN\rangle 
     + \eta^{ij} \langle R(\NN, e_{i})\NN, e_{j}\rangle - \langle \rK^{ij}, \NN\rangle \langle \rK_{ij}, \NN \rangle +\big(\eta^{ij} \langle  \rK_{ij}, \NN \rangle\big)^2\Big{]} \nonumber \\
+    \tfrac{1}{3!} T_{2} \int d^3 V \ \epsilon^{ijk}\Big{[} (3F_{DABC}\NN^{D}(\nabla_{e_{i}}\NN^{A})e_{j}^{B}e_{k}^{C} + (\nabla_{L}F_{ABCD})\NN^{L}\, \NN^D\,  E^{A}(e_{i})E^{B}(e_{j})E^{C}(e_{k}) \Big{]} \ , 
\end{align}
where $R$ is   the Riemann curvature.\foot{When a  membrane  is not coupled to $C_3$ in \rf{032},  \rf{q1}
 it becomes  the equation for a minimal surface in the target space, i.e. 
   $\eta^{ij}\rK_{ij}=0$ and then \rf{5}  follows from a known expression  for the second variation  of the  minimal volume 
  action (see,  e.g., \cite{Simons}).}
   Using   an orthonormal basis $\rn_{p}$    ($p=1, ...,8$)  in  the normal bundle we get 
\begin{align} \label{6a}
 &   S_{B,2} = - T_{2} \int d^3 V \Big{[} \eta^{ij} (\nabla^{\perp}_{e_{i}}\NN)^{p}(\nabla^{\perp}_{e_{j}}\NN)_{p}  +\NN^{p}M_{pq}\NN^{q}\Big{]}  + ( \text{$F_4 $-terms } )\ , 
\\ &M_{pq}=\langle R(\rn_{p}, e_{i})\rn_{q}, e_{j}\rangle\eta^{ij}-\langle \rK^{ij}, \rn_{p}\rangle \langle \rK_{ij}, \rn_{q} \rangle+\langle \text{tr}\,  \rK , n_{p}\rangle \langle \text{tr}\, \rK, \rn_{q} \rangle \ , \ \ \ \text{tr}\, \rK  = \eta^{ij}\rK_{ij} \ .\la{7a}
\end{align}
The quadratic fermionic  part of the M2  brane action in \rf{33}  can be written also as 
(see, e.g.,   \cite{Sakaguchi:2010dg,Beccaria:2023ujc})
\begin{align}
   &\qquad \qquad \qquad  S_{F,2} = T_{2}\int d^3 V \, \eta^{ij} \Bar{\theta} (1-\Gamma)\rho_{i}\cD_{e_{j}}\theta  \ , \label{7b}
\\
    &\rho_{i} = E^{A}(e_{i})\Gamma_{A} \ , \ \ \ \ \ \ \ \{\rho_{i},\rho_{j}\}=2\eta_{ij}\mathds{1}_{32} \ , \qquad 
    \quad \Gamma = \tfrac{1}{3!}\epsilon^{ijk}\rho_{i}\rho_{j}\rho_{k} \ , \ \ \ \Gamma^2=1 \ , \ \ \ \rho^{i}\Gamma=\Gamma\rho^{i} = \tfrac{1}{2}\epsilon^{ijk}\rho_{jk} \ .\la{7c}
\end{align}
To   lowest order the   $\kappa$-symmetry 
  acts as  $\delta\theta = (1+\Gamma)\kappa$.  A 
  convenient choice of the $\ka$-symmetry gauge  is $(1+\Gamma)\theta=0$.

For an  M2 brane   with non-trivial dynamics  only  in the   $ S^{7}/\mathbb{Z}_k$  part of 11d space 
 one can use the induced metric \eqref{m5} in local coordinates in \eqref{m4}
 to define the orthonormal  frame on the world volume as  ($\del_i = \del/\del \xi^i$)
\begin{equation} \label{q2}
    e_{i} = \Big(
        \bc^{-1} \big{(}\partial_{\a}-kA_{\alpha}\partial_2 \big{)}, \ 
        k \partial_{2} \Big)
   \ .
\end{equation}
 Then $C_3$  in \rf{7} does not  contribute to 
   the membrane equations of motion \eqref{q1} which are equivalent to\footnote{To make the connection 
  with \eqref{5a}  explicit,  one may  view the M2 brane  world volume as a 3-surface $\cal M $  in 
    $\mathbb{R}\times \mathbb{C}^4$
   and use that ${\cal M} \subset \mathbb{R}\times S^{7} \subset \mathbb{R}\times \mathbb{C}^4$.}
$\eta^{ij}\rK_{ij} = \eta^{ij}\rK(e_{i}, e_{j}) = 0 .
$
Then \rf{6a}  may  be written  as:
\begin{equation}\label{q3}
    S_{B,2} = - T_{2} \int d^3V \Big{[} \eta^{ij}\langle \nabla^{\perp}_{e_{i}}\NN, \nabla^{\perp}_{e_{j}}\NN\rangle  +\eta^{ij}  \langle R(\NN, e_{i})\NN, e_{j}\rangle - \langle \rK^{ij}, \NN\rangle \langle \rK_{ij}, \NN \rangle \Big{]} \ . 
\end{equation}
Using an orthonormal basis $\rn_{p}$ in the normal bundle we have 
\begin{align}\label{q31}
    &\NN=\NN^{p}\rn_{p} \ , \ \ \ \ \ \ \langle \rn_{p} , \rn_{q} \rangle = \delta_{pq}\ , \ \ \ \ \ \ \ \  \langle \rK_{ij} , \rn_{p} \rangle = \langle \nabla_{e_{i}}e_{j} , \rn_{p} \rangle \ , \\
     &\nabla^{\perp}_{e_i}\NN=\nabla^{\perp}_{e_i}(\rn_{p}\NN^{p}) =\rn_{p}(\partial_{e_i}\NN^{p}) + \rn_{q}\Omega_{ \ p}^{q}(e_{i})\NN^{p} \ , \ \ \  \ \ \  \Omega_{qp}(e_{i}) = \langle \rn_{q}, \nabla_{e_{i}}\rn_{p}\rangle \ .
\end{align}
The fermionic part \eqref{7} is determined by the operator:
\begin{align}\label{q4}
    &\slashed{\cD} = \rho^{i}\cD_{e_{i}} = \rho^{i}\big{[}\nabla_{e_{i}} + \tfrac{1}{12}E^{A}(e_{i})(\Gamma_{A}\rF_{4}-3 \rF_{4A}) \big{]}\ , \ \ \qquad   \nabla_{e_{i}}=\partial_{e_{i}} + \tfrac{1}{4} \Omega^{AB}(e_{i})\Gamma_{AB} \ , \\
    &\rF_{4} \equiv  \tfrac{1}{4!}F_{ABCD}\Gamma^{ABCD} \ , \ \ \qquad   \rF_{4A} \equiv  \tfrac{1}{3!}F_{ABCD}\Gamma^{BCD} \ .
\end{align}
Here  $ \Omega^{AB}$ is the spin connection on \adssZ, $F_{4}=dC_3$  is proportional to the volume form of AdS$_4$
 and $E^{A}$ is a  coframe in \adssZ. 
 Using  the orthonormal
frame $(e_{i}, \rn_{p})$ 
one may  split the $\Gamma^A$-matrices as $(\rho_{i}, \gamma_{p})$.

For  the metric in  explicit coordinates in 
\rf{A12}  we may determine $(e_{i}, \rn_{p})$  in terms of  the local coordinate basis in \adssZ  as follows. 
 For the \sho  membrane solution  corresponding to \rf{314},\rf{m4}    we get (we set $m=1$)   
\begin{align}\label{q5}
    & e_{0} =\bc^{-1} \big{(}\kappa \partial_{t} + ({2}-4a^2)\partial_{\phi_{1}} +4a^2\partial_{\phi_{2}} \big{)} \ , \ \ \ 
    \ \ \ \ \ e_{1} =\bc^{-1}   \big{(}\partial_{\psi} - k(2a^2-\tfrac{1}{2})\partial_{\y}\big{)} \ , \ \ \ e_{2}=k\partial_{\y} \ ,  \nonumber \\
    &\rn_{i} = \partial_{\eta^i} \ , \ \ \ \rn_{4}=\partial_{\chi} \ , \ \ \ \rn_{5} ={\sqrt{2}}{a}^{-1} \partial_{\theta_{1}} \ , \ \ \ \rn_{6}={\sqrt{2}}{({\ha -a^2})^{-1/2}}\partial_{\theta_{2}} \ , \nonumber \\
   & \rn_{7} = 2(\partial_{\phi_1} - \partial_{\phi_{2}}) \ , \ \ \ \ \ \ \  \rn_{8} = 2 \partial_{t}+2 {\sqrt{1-2a^2}}\, {a}^{-1} \partial_{\phi_{1}}+{4a}({{1-2a^2}})^{-1/2}\partial_{\phi_{2}}  \ ,
\end{align}
where $\kappa = 4\sqrt{2}a\sqrt{\ha - a^2}$ and $\bc =2a\sqrt{\ha -a^2}$.\foot{Here $\rn_{i}=\partial_{\eta_{i}}$ ($i=1,2,3$)  correspond  to the normal directions in AdS$_4 $
where $\eta^i$   are the   ``Cartesian" part of  coordinates in AdS$_4$,  i.e. $
ds^2_{\rm AdS_4} = - { (1 + \eta^2)^2 \ov( 1-\eta^2)^2}  dt^2  + { 4 \ov (1-\eta^2)^2}  d \eta^i d \eta^i$.}
 For the \lon  membrane solution with $m=1$  corresponding to \rf{322},\rf{324} we get 
\begin{align}\label{q6}
    & e_{0} =\bc^{-1} \big{(}\kappa \partial_{t} + 2\mathcal{J}(\partial_{\phi_{1}}+\partial_{\phi_{2}}) \big{)} \ ,\ \  \ \ \ e_{1} = \bc^{-1} \partial_{\psi} \ , \ \ \ \ \ \ e_{2}=k\partial_{\y} \ ,  \nonumber \\
    &\rn_{i} = \partial_{\eta^i} \ , \ \  \ \ \ \ \rn_{4}=\partial_{\chi} \ , \ \ \ \ \ \ \ \rn_{5} = 2\sqrt{2}\partial_{\theta_{1}} \ , \ \ \ \rn_{6}=2\sqrt{2}\partial_{\theta_{2}} \ , \nonumber \\
   & \rn_{7} = 2(\partial_{\phi_1} - \partial_{\phi_{2}}) \ ,\ \ \ \ \  \ \ \ \rn_{8} = 4 \mathcal{J}\partial_{t}+2\kappa(\partial_{\phi_{1}}+\partial_{\phi_{2}})
    \ ,
\end{align}
where $\kappa = \sqrt{4\mathcal{J}^2+1}$ and $\bc=\frac{1}{2}$.

 In both cases, the non-zero part of the fermionic operator \eqref{q4} takes the form:
\begin{equation}\label{q7}
    \slashed{\cD}=\rho^{i}\cD_{e_{i}} = \rho^{i}\nabla^{\perp}_{e_{i}} - \tfrac{3}{4}(\rn_{8})^{t}\gamma_{8}\gamma_{1}\gamma_{2}\gamma_{3} \ , \qquad \ \ \ \nabla^{\perp}_{e_{i}} = \partial_{e_{i}} + \tfrac{1}{4}\Omega^{pq}(e_{i})\gamma_{pq} \ .
\end{equation}
We  note that the part with $\rho^{i}\rho^{j}\Omega_{jp}(e_{i})$ is zero since 
\begin{equation}
    \rho^{i}\rho^{j}\Omega_{jp}(e_{i}) = \rho^{i}\rho^{j}\langle e_{j}, \nabla_{e_i}\rn_{p} \rangle = -\rho^{i}\rho^{j}\langle  \nabla_{e_i}e_{j}, \rn_{p} \rangle = -\tfrac{1}{2}\{\rho^{i}, \rho^{j}\}\langle \rK_{ij}, n_p \rangle = -(\text{tr}\ \rK)_{p}= 0 \ ,
\end{equation}
and also that the projection of the connection $\nabla^{\text{T}}$ on the M2 brane
 vanishes because the induced metric is flat.

\iffa
[{\bf to be added??? or not? ?????} ]

The connections on the normal bundle $\Omega_{pq}$ and extrinsic curvatures for both "short" and "long" membranes are presented in the Appendix 
. Here we note that in both cases $\Omega_{pq}$ vanishes in \ads \ normal directions $\rn_{1-3}$. Thus, $\gamma_{1}\gamma_{2}$ commutes with the fermionic operator \eqref{q7} and, at the same time, it commutes with $\Gamma$ defined in \eqref{7c} and used to fix the $\kappa$-symmetry gauge. Therefore, the matrix $\gamma_{1}\gamma_{2}$ can be used to project on its eigenspaces corresponding to the eigenvalues $\pm i$ to simplify the eigenmodes equation discussed in the following section.
\fi

\renewcommand{\theequation}{B.\arabic{equation}}
 \setcounter{equation}{0}
\section{\la{ap2}   Fluctuation frequency    polynomials for   \sho M2  brane}

 The determinant of  the  $8\times 8$   bosonic  fluctuation  operator (with $m=1$)
  in the Fourier representation  \rf{332} that appears in \rf{sm2} 
     may be written as (cf. \rf{3322},\rf{3311},\rf{354}) 
\begin{align} \la{b1}
&\qquad \OD_{\rm B} 
 = (-\omega ^2 + 4 \mathcal{J}+p^2 )^3\,  P_{\rm B}(\omega, n, l,  \tau,  \mathcal{J})
\ ,\\ &
p^2= (n - \tau_1 l)^2 + ( \tau_2 l)^2 \ , \qquad 
q = n - \tau_1 l \ , \ \ \ \  \tau_1= - \ha k \sqrt{1-2 \J} , \qquad   \tau_2 = \tfrac{1}{\sqrt 2}  k \sqrt{\J} \ . \la{b2}
\end{align}
Here $P_{\rm B}$   is an order 10  polynomial in $\omega$
\begin{align} 
P_{\rm B}= &\ \ 
\omega ^{10}+\omega ^8 (12 {\J}-5 p^2-8)\no\\
&+2 \omega ^6 \Big[18 {\J}^2-4 {\J} (5 p^2-2 q^2+6)-4 \sqrt{2-4 {\J}} \sqrt{{\J}} q \sqrt{p^2-q^2}+5 p^4+12 p^2-4 q^2+8\Big]\no\\
& +\omega ^4 \Big[24 \sqrt{2-4 {\J}} {\J}^{3/2} q \sqrt{p^2-q^2}+32 {\J}^3-4 {\J}^2 (11 p^2+12 q^2+4)+8 \sqrt{{\J}} \sqrt{2-4 {\J}} (3 p^2-4) q \sqrt{p^2-q^2}
\no \\
&\qquad +8 {\J} [6 p^4+p^2 (7-6 q^2)+11 q^2]-2 [5 p^6+12 p^4+p^2 (8-12 q^2)+16 q^2]\Big]\no\\
&+\omega ^2 \Big[
-16 \sqrt{2-4 {\J}} {\J}^{3/2} q \sqrt{p^2-q^2} (3 p^2+2 q^2-2)-16 \sqrt{2-4 {\J}} {\J}^{5/2} 
q \sqrt{p^2-q^2}+32 {\J}^3 \left(q^2-2 p^2\right) \la{b3}\\\no
&\qquad +4 {\J}^2 [-9 p^4+8 p^2 \left(2 q^2+3\right)+4 q^2 \left(4 q^2-5\right)]-8 \sqrt{2-4 {\J}} \sqrt{{\J}} p^2 \left(3 p^2-8\right) q \sqrt{p^2-q^2}\\\no
 &\qquad-8 {\J} [3 p^6-6 p^4 \left(q^2+1\right)
+p^2 \left(22 q^2+4\right)+2 q^2 \left(q^2-2\right)]+5 p^8+8 p^6-8 p^4 \left(3 q^2+2\right)+64 p^2 q^2\Big]\no\\
&
-(6 {\J}+p^2-4) \Big[16 \sqrt{2} {\J}^{5/2} q \left(4 q^2-3 p^2\right) \sqrt{p^2-q^2}+8 \sqrt{2} {\J}^{3/2} q \left(2 p^4+3 p^2-4 q^2\right) \sqrt{p^2-q^2}\no\\
&\qquad\qquad \qquad\qquad +16 {\J}^2 \left(p^4-5 p^2 q^2+4 q^4\right)-8 \sqrt{2} \sqrt{{\J}} p^4 q \sqrt{p^2-q^2}\no\\
&\qquad\qquad \qquad\qquad -2 {\J} \left[5 p^6+p^4 \left(4-8 q^2\right)-12 p^2 q^2+8 q^4\right]+p^8+4 p^6-8 p^4 q^2\Big]\ . \no 
\end{align}
For the  fermionic  fluctuations   we find
\be \la{b5}
\OD_{\rm F}  = \big[P_{{\rm F}}(\omega, n,l,\tau, \mathcal{J})\big]^2  \ , 
\ee
where $P_{\rm F}$  is 4-th order polynomial in $\omega^2$  
\begin{align} 
     P_{\rm F} =&\omega ^8-4 \left(p^2+1\right) \omega ^6 +
     \omega ^4 \Big[6 J^2+4 J \left(p^2+2 q^2-2\right)-4 \sqrt{2-4 J} \sqrt{J} q \sqrt{p^2-q^2}+6 p^4+8 p^2-4 q^2+6\Big]\no\\
  &   +\omega ^2 \Big[
  16 \sqrt{2-4 J} J^{3/2} q \sqrt{p^2-q^2}-16 J^3+4 J^2 \left(3 p^2-8 q^2-3\right)+8 \sqrt{J} \sqrt{2-4 J} \left(p^2+1\right) q \sqrt{p^2-q^2}\no\\
  & \qquad\qquad -8 J [p^4+p^2 \left(2 q^2+1\right)-2]
  -4 \left(p^2+1\right) \left(p^4-2 q^2+1\right)     \Big] \la{b6}
  \\
&-8 \sqrt{2-4 J} J^{3/2} q \sqrt{p^2-q^2} \left(3 p^2+q^2-4\right)-60 \sqrt{2-4 J} J^{5/2} q \sqrt{p^2-q^2}+9 J^4-12 J^3 \left(5 p^2-10 q^2+2\right)\no\\
&\qquad +2 J^2 \left[-5 p^4+4 p^2 \left(5 q^2+7\right)+8 q^4-62 q^2+11\right]+4 \sqrt{J} \sqrt{2-4 J} q \left(-p^4+2 q^2-1\right) \sqrt{p^2-q^2}\no\\
&\qquad +4 J \left[p^6+p^4 \left(2 q^2+1\right)-p^2 \left(4 q^2+3\right)-5 q^4+10 q^2-2\right]+\left(p^4-2 q^2+1\right)^2\ .\no
\end{align}

\renewcommand{\theequation}{C.\arabic{equation}}
 \setcounter{equation}{0}
\section{\la{ap3} Fluctuation frequency    polynomials for   \lon M2 brane } 


Let us start with the bosonic fluctuations. We will specify to  the case of  the  minimal winding 
$m=1$.   The determinant of  the  $8\times 8$    fluctuation  operator 
 in the Fourier representation  \rf{332} that appears in \rf{sm2}    may be written as (cf. \rf{3322}) 
\be \la{c1}
\OD_{\rm B}  = (-\omega ^2 + 4 \mathcal{J}^2+n^2+\qq^2 + 1 )^3\,  P_{\rm B}(\omega, n, \qq,  \mathcal{J}) \ , \ \ \ \ \ \ \ \ \ \ 
\qq\equiv  \tau_2 l = \ha k l  \ ,  \ee
where $P_{\rm B}$  given by 
\begin{align}
&P_{\rm B}=  \det \begin{pmatrix}
n^2 + \qq^2 - \omega^2 - 1 & 0 & 0 & 2i\mathcal{J}\omega & 0 \\
0 & 4\mathcal{J}^2 + n^2 + \qq^2 - \omega^2 & 0 & -\frac{i(n+\qq)}{\sqrt{2}} & -\frac{1}{2}i\sqrt{8\mathcal{J}^2 + 2}(n+\qq) \\
0 & 0 & 4\mathcal{J}^2 + n^2 + \qq^2 - \omega^2 & -\frac{i(n-\qq)}{\sqrt{2}} & \frac{1}{2}i\sqrt{8\mathcal{J}^2 + 2}(n-\qq) \\
-2i\mathcal{J}\omega & \frac{i(n+\qq)}{\sqrt{2}} & \frac{i(n-\qq)}{\sqrt{2}} & n^2 + \qq^2 - \omega^2 & 0 \\
0 & \frac{1}{2}i\sqrt{8\mathcal{J}^2 + 2}(n+\qq) & -\frac{1}{2}i\sqrt{8\mathcal{J}^2 + 2}(n-\qq) & 0 & n^2 + \qq^2 - \omega^2
\end{pmatrix} 
\end{align}
It is a  polynomial of order $5$ in $\omega^2$  with the explicit form being 
\begin{align} \label{B4}
    & P_{\rm B}
     =\omega ^{10} + \omega ^8 (-12 J^2-5 n^2-5 \qq^2+1)+\omega ^6 \big[48 J^4+8 J^2 (5 n^2+5 \qq^2-1)+2 (n^2+\qq^2) (5 n^2+5 \qq^2-3)\big] \nonumber \\
    &\qquad -2 \omega ^4 \big[32 J^6+8 J^4 (5 n^2+5 \qq^2-1)+8 J^2 (n^2+\qq^2) (3 n^2+3 \qq^2-2)+(n^2+\qq^2)^2 (5 n^2+5 \qq^2-6)+n^2+\qq^2) \big]\nonumber \\
    &\qquad \ \ +\omega ^2 \Big[ 32 J^4 (n^2+\qq^2-1) (n^2+\qq^2)+8 J^2 \big[ 3 n^6+n^4 (9 \qq^2-4)+n^2 (9 \qq^4-10 \qq^2+1)+3 \qq^6-4 \qq^4+\qq^2\big] \nonumber \\
    & \qquad\qquad \ \ \ \  +5 n^8+10 n^6 (2 \qq^2-1)+5 n^4 (6 \qq^4-6 qq^2+1)+n^2 (20 \qq^6-30 \qq^4+6 \qq^2)+5 \qq^4 (\qq^2-1)^2\Big] \nonumber \\    
    &\qquad \quad -(n^2+\qq^2-1)  \Big[n^6 (4 J^2+4 \qq^2-2)+n^4 [4 J^2 (3 \qq^2-1)+6 \qq^4-6 \qq^2+1] \nonumber \\
    & \qquad \qquad  \qquad \qquad\qquad +2 n^2 \qq^2 \big[6 J^2 (\qq^2-2)+2 \qq^4-3 \qq^2-1\big]+\qq^4 (\qq^2-1) (4 J^2+\qq^2-1)+n^8\Big] \ .
\end{align}
The characteristic frequencies are solutions of $\OD_{\rm B} =0$. 
The 3   decoupled  modes  with 
\begin{equation}
    \omega = \sqrt{4\mathcal{J}^2+n^2+\qq^2+1} \ .
\end{equation}
correspond to the transverse fluctuations of the M2 brane in the \ads\ directions.   

\iffa 
Note that if we  formally expand   the frequencies at large $\J$ as 
\begin{equation} \label{B3}
    \omega = v_{1}\mathcal{J} + \frac{v_{2}}{\mathcal{J}} + \mathcal{O}\big{(}\frac{1}{\mathcal{J}^3}\big{)} \ , 
\end{equation}
we find  (assuming $\omega >0$) 
that there are two modes with $v_{1}=0$ and 
\begin{equation}
    v_2 = \frac{1}{2}\sqrt{(n^2+\qq^2)^2-(n^2+\qq^2)\pm 2\sqrt{n^2\qq^2(n^2+\qq^2-1)}}  \ ,
\end{equation}
which in the string case  of $l=0$ or $\qq=0$ correspond to the expansion
of  (cf. \rf{lm1}) 
\begin{equation}
    \omega=  \sqrt{n^2 + 2\mathcal{J}^2 -\sqrt{4\mathcal{J}^4+n^2(4\mathcal{J}^2+1)}} =  \frac{\sqrt{n^2(n^2-1)}}{2\mathcal{J}}+\mathcal{O}\big{(}\frac{1}{\mathcal{J}^3}\big{)} \ .\la{c4}
\end{equation}
\fi
\iffa 
The other three modes have $a_{1}=2$, for which $a_{3}$ coefficients are determined by the cubic equation:
\begin{equation}
    a_{3}-\frac{1}{4}a_{3}^2(5p^2-1)+\frac{1}{2}a_{3}(p^4-\frac{1}{2}p^2)-\frac{1}{16}(p^6-p^4+n^2\qq^2) = 0 \ ,
\end{equation}
where $p^2=n^2+\qq^2$. Although one can try to find the solutions for $a_{3}$ explicitly, we are interested in their sum given by:
\begin{equation}
    a_{3}^{(1)}+a_{3}^{(2)}+a_{3}^{(3)} = \frac{1}{4}(5(n^2+\qq^2)-1) \ .
\end{equation}
These modes in the case $q=0$ correspond to the sum of the remaining frequencies given in \eqref{12A}, \eqref{12B}. Thus, we arrive at the following expansion of $\Omega_{B}(n, q; \mathcal{J})$:
\begin{align}\label{B1}
    &\Omega_{B}(n, q;\mathcal{J}) = 12\mathcal{J} + \frac{1}{2\mathcal{J}}\big{(}4(n^2+\qq^2)+1  \nonumber \\
    &+ \sqrt{(n^2+\qq^2)^2-(n^2+\qq^2)+ 2\sqrt{n^2\qq^2(n^2+\qq^2-1)}}+\sqrt{(n^2+\qq^2)^2-(n^2+\qq^2)- 2\sqrt{n^2\qq^2(n^2+\qq^2-1)}}\big{)} +\mathcal{O}\big{(}\frac{1}{\mathcal{J}^3}\big{)}\ .
\end{align}
\fi

Similarly, for the  fermionic  fluctuations   we find
\be \la{c5}
\OD_{\rm F}  =  \big[P_{{\rm F}}(\omega, n, \qq, \mathcal{J})\big]^2  \ , 
\ee
where $P_{\rm F}$  is 4-th order polynomial in $\omega^2$  
\begin{align} \label{B5}
     P_{\rm F} =&\omega ^8+\omega ^6 (-12 J^2-4 n^2-4 \qq^2-1)+\omega ^4 \big[30 J^4+J^2 (32 n^2+32 \qq^2+5)+6 (n^2+\qq^2)^2+2 n^2+2 \qq^2+\tfrac{3}{8}\big] \nonumber \\
    &\quad  -\omega ^2 (4 J^2+4 n^2+4 \qq^2+1) \big[7 J^4+6 J^2 (n^2+\qq^2)+(n^2+\qq^2)^2+\tfrac{1}{16}  \big] \nonumber \\
    &+\qq^8  + 4 \qq^6 (2 J^2+n^2) + \qq^4 \big[22 J^4+2 J^2 (24 n^2+1)+6 n^4-\tfrac{1}{8}\big]   \\
    &+\tfrac{1}{4} \qq^2 \big[ 96 J^6+16 J^4 (11 n^2+1)+J^2 (96 n^4+24 n^2-2)+16 n^6+3 n^2\big] 
    +\tfrac{1}{256} (4 J^2+4 n^2+1)^2 (12 J^2+4 n^2-1)^2\no \quad 
\end{align}
Thus  each $\omega$ that is solves  $\OD_{\rm F} =0$ 
 has degeneracy  two.

\iffa 
Expanding in large $\J$ as in \rf{B3}  we get two positive roots with $v_{1}=3$:
\begin{equation}
    \omega = 3\mathcal{J} +\frac{1}{8\mathcal{J}}\big[1+2(n^2+\qq^2)\big] +\mathcal{O}\big(\frac{1}{\mathcal{J}^3}\big)\ , 
\end{equation}
which for $l=0$ or  $\qq=0$ correspond to the expansion of (cf. \rf{lm2})  {[\bf which modes ???]} 
\begin{equation}
    \sqrt{4\mathcal{J}^2+n^2}+\frac{1}{2}\sqrt{4\mathcal{J}^2+1} = 3 \mathcal{J}+\frac{2n^2+1}{8\mathcal{J}}+\mathcal{O}\big{(}\frac{1}{\mathcal{J}^3}\big{)} \times 2  \ .
\end{equation}
\fi

\renewcommand{\theequation}{D.\arabic{equation}}
 \setcounter{equation}{0}
\section{\la{ap4}  Non-planar corrections to  ABJM \brf}
In the case of the  $\N=4$  $SU(N)$ SYM  theory  the \brf  may be found  from the exact localization result \ci{Drukker:2000rr}   for the expectation value of the $\ha$ BPS circular Wilson loop as 
  \ci{Correa:2012at}
 \begin{align} 
 &\qquad \qquad \qquad  B_{_{\rm SYM}}= {1\ov 2 \pi^2} \l {\del \ov \del  \l} \log \langle W \rangle_{_{\rm SYM}} \ ,\la{b101} \\
 & \qquad 
 \langle W \rangle_{_{\rm SYM}} = 
 e^{{\l\ov 8 N^2}(N-1)} L^1_{N-1} ( - {\l\ov 4 N})
  = {2 N\ov \sql} I_1(\sql) \Big(1 + {{ \l^{3/2}} \ov 96 N^2  } \Big[  { I_2(\sql) \ov I_1(\sql)} - {12 \ov \sql} \Big]  + ... \Big)  \ , \la{b11} \\
 &\la{bb12}
 B_{_{\rm SYM}}(\l, N) 
= B^{(\infty)}_{_{\sym}}(\l) +  {1\ov 128 \pi^2} { \l^{3/2} \ov N^2} + ...    \ , \qquad \qquad 
B^{(\infty)}_{_{\sym}}(\l) = {\sql \ov 4 \pi^2} { I_2(\sql)\ov I_1(\sql)} 
={\sql\ov 4 \pi^2} - {3\ov 8 \pi^2} +....\ . \end{align}
 To get the  \brf  one may  use  the original definition as a derivative over the angle  of a small cusp  or 
 one  may     start with the expression for the $\ha$ BPS Wilson loop 
 wrapped $\rr$ times on the circle and then  \ci{Lewkowycz:2013laa}
 \be \la{b51}
B (\l,N)   = {1\ov 4 \pi^2}  {\del\ov \del \rr} \log\langle  W\rangle \Big|_{\rr=1} \ . 
\ee
In the $\N=4$ SYM case this leads to the same expression as in \rf{b11}
since the dependence on $\rw$ can be incorporated into  $ \langle  W\rangle$ in \rf{b11}  by $ \sql \to \rw \sql$.

Ref. \ci{Lewkowycz:2013laa}  has shown that \rf{b51} applies also in the ABJM case  for 
the \brf given   in terms  of the  $1\ov 6$ BPS Wilson loop defined on a small cusp. 
One  may  also  find  the  \brf   corresponding to either  ${1\ov 2}$ or $1\ov 6$  BPS Wilson loops
by using a  generalization of the  identity  \ci{Correa:2012at}
that expresses $B (\l,N) $    as a 
 derivative of  the  logarithm of the  latitude Wilson loop  with respect to the small
   latitude angle \ci{Bianchi:2014laa,Correa:2014aga,Bianchi:2017ozk,Bianchi:2018scb}.\foot{For ${1\ov 2}$   BPS Wilson loop this identity was proposed and proved perturbatively in \ci{Bianchi:2014laa}, 
and for the corresponding \brf  it  was first introduced   and then proved exactly in \ci{Bianchi:2017ozk}.
In  the ${1\ov 6}$   BPS Wilson loop  case a similar identity for the  \brf  was proved in \ci{Correa:2014aga} and further 
elaborated on in  \ci{Bianchi:2018scb}.
For a
 review of   
  the 
  \brf   
 in the ABJM theory   see  the contribution of L. Bianchi in \ci{Drukker:2019bev}   and  also \ci{Penati:2021tfj}.} 
In the planar ($N=\infty$)  limit  one finds 
the following  strong coupling expansion  for the \brf   corresponding to the ${1\ov 2}$  BPS  Wilson loop
 \ci{Bianchi:2018scb}
\be \la{b4}
B^{(\infty)}_{\abjm} = {1\ov 2\pi} \sqrt{ \l\ov 2} - { 1\ov 4 \pi^2} - {1\ov 96 \pi} {1\ov \sqrt{2 \l}}  + ... \ ,  \ee 
which matches  the string theory prediction  at the two leading orders   \ci{Forini:2012bb,Aguilera-Damia:2014bqa}. 
Finding 
non-planar corrections  
  in this approach is  hard as the exact  localization result  is  not known for a non-trivial cusp angle. 
  An alternative approach was suggested in 
 \ci{Guerrini:2023rdw,Armanini:2024kww}.

 One may conjecture that  in general the multi-wrapped Wilson loop expectation  value   is the same as 
 the one  for the  loop   in the $\rw$-fundamental representation. 
 The corresponding localization result was  found 
  in    \cite{Klemm:2012ii} 
   and is a simple generalization  of the expression  given above in \rf{2190}
\begin{equation}
\langle W \rangle_{_{\rm ABJM}} = \frac{1}{2\, \sin \frac{2\pi\rr}{k} } \frac{{\rm Ai}\left[( {\pi^2\ov 2} k)^{{1}/{3}}\left(N-\frac{k}{24}-\frac{1}{3k }- \frac{2\rr}{k}\right)\right]}{{\rm Ai}\left[( {\pi^2\ov 2} k)^{{1}/{3}}\left(N-\frac{k}{24}-\frac{1}{3k}\right)\right]}\ . 
\label{2107a}
\end{equation}
We have checked explicitly that  using  \rf{2107a} in \rf{b51}  one  gets  the result 
for  the corresponding \brf   which is equivalent to  the one  found using other  definitions of  $B (\l,N)$
in the  $\ha$  BPS  Wilson loop case 
 \ci{Bianchi:2018scb,Bianchi:2018bke,Armanini:2024kww} (see eq. (7.14) in \ci{Armanini:2024kww}).

The   expansion  of \rf{2107a}  at  large $N$  for  fixed  $k$ 
is similar to the one in \rf{350}
\begin{equation}
\langle W\rangle_{_{\rm ABJM}} = \frac{1}{2\sin\frac{2\pi\rr}{k}}\, e^{\pi \rr \sqrt{\frac{2N}{k}}}\Big[
1-\frac{\pi \rr \left(k^2+24\rr +8 \right)}{24 \sqrt{2} \, k^{3/2}}\frac{1}{\sqrt{N}}+O(\frac{1}{N})
\Big] \ . 
\label{220a}
\end{equation}
Using this in \rf{b51} 
 and  expanding  in  large $k$  we then get 
\begin{align}
 B_{\abjm  }  &= {1\ov 4 \pi} \sqrt{2 N\ov k}   - {1\ov 2 \pi k } \cot{ 2 \pi\ov k} 
+ ... 
= {1\ov 2\pi} \sqrt{ \l\ov 2} 
-\frac{1}{4 \pi ^2} +... + \frac{1}{3 k^2} +   \frac{4 \pi ^2}{45 k^4}   +\frac{32 \pi ^4}{945 k^6}
 +  ... \no \\ 
 &= \la{b8}
B^{(\infty)}_{\abjm} +\frac{\l^2}{3 N^2} +  \frac{4 \pi ^2\l^4}{45 N^4}  +\frac{32 \pi ^4\l^6 }{945 N^6} + ... \ . 
\end{align}
It is interesting to note that 
 the coefficients of the first two  leading    non-planar corrections here are the same
 (up  to  an  overall $2\pi$  factor)  as 
 in  the cusp anomaly function  in \rf{777},\rf{211}.     

It would be important  to explain the dependence of \rf{220a} on $\rr$   from 
 the semiclassical M2 brane point of view.
 One  possible  approach is   to generalize the discussion in  \ci{Giombi:2023vzu} to the case  when the 
 minimal surface is wrapped  $\rr$ times  on the  boundary circle.  
 While  the dependence of exponential in \rf{220a}  on $\rr$ then  follows simply 
  from the value of the classical 
 action,  it is not clear how a   particular   $\rr$-dependence  of the  $1\ov \sin$  prefactor in \rf{2107a},\rf{220a} 
   may  come out of the 1-loop M2  brane   contribution    generalizing $\rr=1$    one in 
   \ci{Giombi:2023vzu}. 
   
   Somewhat  surprisingly,  the dependence of the  tree-level $e^{\pi \rr \sqrt{\frac{2N}{k}}}$ and 1-loop $\frac{1}{2\sin\frac{2\pi\rr}{k}}\, $  
   prefactors 
    in \rf{220a} on $\rr$ is actually 
    the same 
    as in the   case  when the  M2  brane  is  wrapped $\rr$ times   not on the AdS$_4$ boundary 
    circle   but on the 11d circle $\vp$.
     In this case we   have effectively 
    $\vp \to \rr \vp$   and thus the radius $1/k$   in \rf{06}  is    rescaled to $\rr/k$. 
    This  leads  to  $\frac{2\pi }{k} \to \frac{2\pi\rr}{k}$ in the M2 brane 1-loop correction.
The factor of $\rr$ in the exponent in \rf{220a}  then also has  an obvious  origin: 
the classical M2 brane action  is proportional to  the length of the 11d circle, i.e. 
$ { 2 \pi \rr\ov k}  $,   with the additional dependence on $N$ and $k$  coming from 
the effective M2  brane tension  factor $\T_2 $ in \rf{011}. 
However, the  $\rr$-dependence of   the subleading  terms in \rf{220a}  (that should originate from the  two 
 and higher loop M2  brane corrections)  does not appear to have a similar simple explanation. 
 

 \np
\small 


\ed